\documentclass[onecolumn,na4paper, amsfonts, amssymb, amsmath, reprint, showkeys, nofootinbib, twoside,superscriptaddress]{revtex4-1}
\usepackage[english]{babel}
\usepackage[utf8]{inputenc}
\usepackage{bm}
\usepackage{comment,ifthen}
\usepackage[normalem]{ulem}
\usepackage[colorinlistoftodos, color=green!40, prependcaption]{todonotes}
\usepackage[pdftex, pdftitle={Article}, pdfauthor={Author}]{hyperref} 

\newcommand{\shcm}{0}
\ifthenelse{\equal{\shcm}{1}}{
\newcommand{\tds}[1]{\textcolor{olive}{\bf TDS: #1}}
\newcommand{\pg}[1]{\textcolor{purple}{\bf PG: #1}}
\newcommand{\mcm}[1]{\textcolor{teal}{\bf MCM: #1}}
\newcommand{\jrk}[1]{\textcolor{green}{\bf JRK: #1}}
\newcommand{\change}[1]{\textcolor{red}{\bf #1}}
\newcommand{\ag}[1]{\textcolor{blue}{\bf AG: #1}}
}{
\newcommand{\tds}[1]{}
\newcommand{\pg}[1]{}
\newcommand{\mcm}[1]{}
\newcommand{\jrk}[1]{}
\newcommand{\ag}[1]{}
\newcommand{\change}[1]{{#1}}
}

\begin{document}
\title{Synergistic coupling in \textit{ab initio}-machine learning simulations of dislocations}

\author{Petr Grigorev}
\email{grigorev@cinam.univ-mrs.fr}
\affiliation{Aix-Marseille Universit\'{e}, CNRS, CINaM UMR 7325, Campus de Luminy, 13288 Marseille, France}

\author{Alexandra M. Goryaeva}
\address{Universit\'e Paris-Saclay, CEA, Service de Recherches de M\'etallurgie Physique, 91191, Gif-sur-Yvette, France}%

\author{Mihai-Cosmin Marinica}
\address{Universit\'e Paris-Saclay, CEA, Service de Recherches de M\'etallurgie Physique, 91191, Gif-sur-Yvette, France}%

\author{James R. Kermode}
\address{Warwick Centre for Predictive Modelling, School of Engineering, University of Warwick, Coventry CV4 7AL, United Kingdom}

\author{Thomas D Swinburne}
\email{swinburne@cinam.univ-mrs.fr}
\affiliation{Aix-Marseille Universit\'{e}, CNRS, CINaM UMR 7325, Campus de Luminy, 13288 Marseille, France}
\date{\today} 

\begin{abstract}
\textit{Ab initio} simulations of dislocations are essential to build quantitative models of material strength, but the required system sizes are often at or beyond the limit of existing methods. Many important structures are thus missing in the training or validation of interatomic potentials, whilst studies of dislocation-defect interactions must mitigate the effect of strong periodic image interactions along the line direction.
We show how these restrictions can be lifted through the use of linear machine learning potentials in hybrid simulations, where only a subset of atoms are governed by \textit{ab initio} forces. The linear form is exploited in a constrained retraining procedure, qualitatively expanding the range of 
training structures for learning and giving precise matching of dislocation core structures, such that lines can cross the quantum/classical boundary.
We apply our method to fully three dimensional studies of impurity segregation to edge and screw dislocations in tungsten. Our retrained potentials give systematically improved accuracy to QM/ML reference data and the three dimensional geometry allows for long-range relaxations that qualitatively change impurity-induced core reconstructions compared to simulations using short periodic supercells. More generally, the ability to treat arbitrary sub-regions of large scale simulations with \textit{ab initio} accuracy opens a vast range of previously inaccessible extended defects to quantitative investigation.
\end{abstract}

\maketitle

\section{Introduction}
    Dislocations are extended line defects which carry plastic deformation in crystalline 
    materials \cite{Hirth}. Alongside twins and grain boundaries, the formation and 
    migration of dislocations are responsible for the ductility, or formability, of metal components - their ability to smoothly deform rather than fracture when loaded beyond their elastic limit.
    Understanding and optimizing dislocation behaviour is a central topic in computational metallurgy \cite{leyson2010quantitative,Rodney_2017}.
    
    The long range elastic fields of dislocations bias the diffusion of point defects such as 
    vacancies or impurity atoms, whose segregation to the highly deformed dislocation `core' can
    qualitatively change the core structure and raise or reduce the intrinsic lattice 
    resistance to dislocation migration \cite{Hu2017}. A classic example, known empirically for 
    thousands of years \cite{schmidt1978complex}, is the pinning of dislocations in iron alloys 
    due to the segregation of interstitial carbon \cite{Ventelon_2015}, giving a harder but also 
    more brittle steel \cite{cottrell1967introduction,argon2008strengthening}. Similar mechanisms 
    play a central role in the solution strengthening of e.g. titanium \cite{Yu2015}, aluminium alloys \cite{Leyson2010} and high entropy alloys \cite{Varvenne2017, Nag2020}.
    Point defects formed under extreme conditions such as neutron irradiation \cite{Zinkle2013, Zheng_VHe_hardening} 
    or plasma exposure \cite{Li2020} similarly affect dislocation motion, being a primary source of
    irradiation-induced embrittlement \cite{swinburne2018kink} \textit{i.e.} a marked reduction in 
    the amount of absorbed mechanical energy before failure \cite{Ritchie2011}. 
    
    A quantitative and mechanistic understanding of point-defect dislocation interactions 
    is therefore of primary importance to build accurate models for rational alloy design 
    strategies \cite{Varvenne2017}, or to predict and mitigate the risk of catastrophic 
    brittle failure during service under hostile conditions \cite{arakawa2021perspectives}.\\
    
    For this task, \textit{ab initio} calculations, specifically density functional theory
    (DFT) \cite{martin}, are essential to capture dislocation core structures, with complex bonding
    to impurity elements. However, the computational cost of DFT \change{typically scales as $\mathcal{O}(N^3)$ for metallic systems, which  limits its direct applicability to the study of extended defects.}
    In certain special cases,
    when the core is sufficiently compact and the elastic field sufficiently weak, it is possible
    to contain specialised dislocation multipole configurations in small periodic DFT supercells \cite{Woodward_2005, Ventelon2007, Rodney_2017}. Whilst simulations of this kind have
    given significant insight into the nature of intrinsic lattice resistance in a variety of
    systems \cite{dezerald2015first,Dezerald2016,clouet2015dislocation}, extending the same
    methodology to dislocations with large prismatic components 
    \pg{Do we mean dissociation into partials here? later in the paper we use term split core with a different meaning}\tds{good point, I'll remove!} is extremely challenging due to the system sizes required to mitigate \change{the resulting} strong elastic interactions.
    The requirement of periodicity along the line direction makes studies of
    impurity segregation similarly expensive, even with amenable dislocations,
    due to long ranged core reconstructions as we show below. Recent studies
    have simulated well over a thousand DFT atoms to infer the
    converged structure \cite{HACHET2020481}.\\
    
    An alternative to fully periodic supercells is a cluster approach \cite{Woodward_2002,Woodward_2005},
    where periodicity is kept only along the line of the dislocation (or other extended defects, such as a crack \cite{Kermode2008})
    with free boundary conditions used in other directions (Methods). In flexible boundary methods the DFT 
    cluster is coupled to a continuum medium via lattice Greens functions \cite{Woodward_2002, Woodward_2005, Trinkle_2018, leyson2010quantitative}. 
    In the hybrid or QM/ML methods employed here, the cluster is coupled to an 
    atomic system governed by an interatomic potential \cite{Kermode_Swinburne_2017,bernstein2009hybrid}.
    Providing the atomic environment at the cluster boundary is described identically by
    both methods, the relaxed configuration is equivalent to an  \change{unfeasibly} large
    \textit{ab initio} simulation. This dictates that deformations must be within
    the elastic regime for flexible boundary methods. Whilst hybrid simulations could have arbitrary
    deformations in principle, in practice empirical interatomic potentials can only be trusted
    in the elastic regime, though it is still common to perform an \textit{ad hoc} rescaling to ensure
    seamless coupling to \textit{ab initio} \cite{Kermode_Swinburne_2017}. Whilst a total energy cannot be 
    rigorously defined in cluster approaches, access to the ionic forces allows energy differences
    to be calculated via the principle of virtual work, which have been validated 
    against total energy calculations in previous works \cite{Kermode_Swinburne_2017,PhysRevMaterials.4.023601} (Methods).\\
    
    Although cluster approaches allow for a much wider range of core structures, the limitation 
    of elastic matching at the cluster boundary still imposes periodicity along the dislocation 
    line direction. As a result, dislocation-impurity studies remain challenging or approximate. 
    As with fully periodic approaches, although such simulations are extremely valuable when possible, 
    even the lower limit of required system sizes are at or beyond the resources of many practitioners, 
    and essentially rule out the systematic studies required for rational design approaches.\\
    
    The \change{severe} size and geometry limitations of DFT are also significant for the development of interatomic potentials, which has been revolutionised by the availability of high-dimensional 
    regression algorithms from the machine learning (ML) community, designed
    to mitigate overfitting issues whilst retaining flexibility \cite{shao1993linear,srivastava2014dropout,mackay1992bayesian}.
    We refer the reader to a number of excellent recent reviews in this rapidly growing field \cite{goryaeva2019towards,deringer2021gaussian,mishin2021machine,onat2020sensitivity, Unke_machine_2021},
    which has attracted explosive interest following the ability of 
    state-of-the-art ML potentials \cite{bartok_thesis, bartok2010,behler2007,Thompson_snap_2015,Shapeev_MTP,Shapeev_MTP2017,goryaeva2021,allen2021atomic,pun2019physically, Chmiela_towards_2018, lysogorskiy2021performant, drautz_atomic_2019, drautz_atomic_2020} \mcm{I patch with many references} to achieve \textit{ab initio} accuracy across a diverse configuration space.
    However, the extrapolation ability of these approaches remains a subject of intense 
    interest \cite{goryaeva2021, Unke_machine_2021}, due in part to the requirement to train and validate 
    only on small, \change{periodic} DFT simulations, whilst the desired applications typically operate on 
    much longer time and length scales. 
    
    For example, a common strategy to model dislocation structures is to train on gamma surfaces,
    inspired by the Peierls-Nabarro model \cite{peierls1940size}.
    Whilst reasonable, this correspondence is only approximate and does not have a direct relevance
    for models aiming to capture more complex mechanisms such as point defect interactions.
    Various learn on-the-fly attempts have been made to overcome this problem
    by continuously extending the training database with configurations encountered in large scale simulations
    and selected by a certain measure of extrapolation \cite{Bernstein_novo_2019, Hodapp2020,vandermause2020fly}.
    Nevertheless, the same fundamental limitations on the range of accessible structures remains.
    
    In this paper, we describe and apply a general method to simulate extended defects in hybrid, or QM/ML,
    simulations employing linear machine learning potentials
    \cite{goryaeva2021}.
    
    Our first main result is a demonstration that `standard' hybrid simulations can be used to retrain
    an existing potential, exploiting the linear form to design a simple refitting procedure that 
    \textit{exactly} preserves desired properties, namely elastic constants. 
    \change{
    We note that similar functionality should be achievable with
    highly flexible (large capacity) approaches employing e.g. neural networks \cite{behler2007,pun2019physically} or kernels \cite{Book_Rasmussen} such as GAP \cite{bartok2010}. Kernel methods can have theoretically infinite capacity, which is controlled during the refitting procedure by sparsification and providing the relevant information from the database  \cite{Bernstein_novo_2019}. However, this procedure requires special design of the database, some prior knowledge of relevant atomic configurations for a specific physical problems pointed up with the appropriate theoretical tools   \cite{Bartok_machine_2017, goryaeva2020, goryaeva2021}.  
    The present procedure is general and simply requires a linear ML potential with capacity larger than the minimum required for the assimilation of elastic properties. Here, we use a quadratic non-linear ML in descriptors, that can be seen as linear ML potential in an extended descriptor space \cite{Thompson_snap_2015, goryaeva2021}. Moreover, recently, it was shown that this formalism has enough learning capacity in order to assimilate complex features of  defects energy landscape in Fe and W  \cite{goryaeva2021}, thus being appropriate for the present investigation.
    }
  %
    This is used to reproduce
    QM/ML simulations of previously unseen core structures and migration barriers
    for prismatic junction dislocations in tungsten, \change{extending} a recently 
    released machine learning potential \cite{goryaeva2021}.\\

    Our second main result is that retrained potentials can extend the range of coupling
    geometries in hybrid simulations, exploiting the exact agreement in dislocation core structure
    to allow dislocations to pierce the hybrid boundary. In place of thousand-atom DFT simulations
    with multiple k-points and periodic image concerns, we present simulations of defect-dislocation
    interaction in tungsten using only $\Gamma$-point calculations with a few hundred atoms,
    namely a spheroid around the impurity or point defect. 
    \jrk{This is OK, but I wouldn't push the $\Gamma$ point line too far as it reduces the scalability of the DFT code} %
    \ag{I agree with James regarding $\Gamma$ points. I would keep it for methods}
    \tds{I would like to keep it in, but not push any harder...}
    \ag{OK}
    Our fully three-dimensional simulations
    involving nearly 100,000 atoms reveal qualitatively distinct core reconstructions compared to
    periodic approaches, with long-range relaxations extending over several nanometers. We also provide a
    validation test of the employed tungsten potential, calculating dislocation-vacancy binding energies.
    An open source-implementation of our method, employing the atomic simulation environment (ASE)
    package \cite{Hjorth_Larsen_2017}, is available online \cite{code}.

\begin{figure*}
    \centering
    \includegraphics[width=\textwidth]{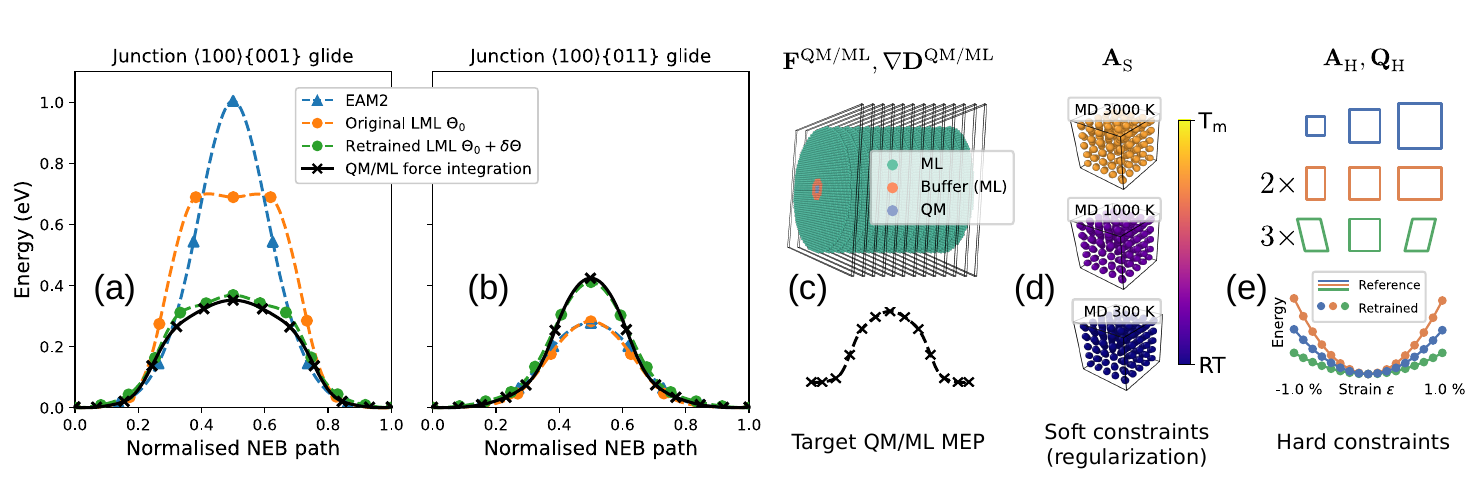}
    \caption{Peierls barriers of edge (or junction) dislocations with $ \bf{b} =$ $\langle100\rangle$ (a) gliding in \{001\} and (b) in \{011\} planes. The barriers are obtained by our NEB method showing the total energy variation along the path for the potentials (dashed lines and dots) and virtual work from force integration for QM/ML results (black solid lines and crosses). (c) An illustration of the QM/ML NEB pathway which the retraining procedure targets. \change{Schematic representations of (d) bulk molecular dynamics snapshots at finite temperatures used to generate `soft' constraints for regularization and (e) strained bulk configurations used to generate `hard' constraints for elastic properties.}
    }
    \label{fig:glide_summary}
\end{figure*}

\section{Results}
\jrk{Feels a bit like we're missing a methodology section. Could II.B be moved from Results to a new Methodology? II.A does belong in Results, so need to check it doesn't break the flow.}
\tds{Not sure if NPJ allow this, but I agree it improves the readability}
\ag{here are the requrements of npj:
The main text (excluding abstract, Methods, references and figure legends) is typically no more than 4,000-4,500 words. The abstract is typically 150 words, unreferenced. Articles have up to 10 display items (figures and/or tables). An Introduction section is followed by sections headed Results, Discussion, Methods and Data Availability. The Results and Methods should be divided by topical subheadings; the Discussion may contain subheadings at the editors' discretion. As a guideline, Articles have around 60 references.}
\jrk{OK, guess it has to go back in results, sorry...}

\subsection{Constrained retraining procedure}
As discussed above, in the present work we employ the linear machine learning (LML)
potential approach \cite{goryaeva2021,Thompson_snap_2015,Shapeev_MTP,allen2021atomic,lysogorskiy2021performant}, which has cohesive energy and gradient forces 
\begin{equation}
    \mathcal{U}^{\rm LML}({\bf X}) = {\bf D}({\bf X})\cdot{\bm\Theta}, 
    \quad
    \mathbf{F}^{\rm LML}({\bf X}) = -{\bm\nabla}\mathbf{D}({\bf X}){\bm\Theta}
    \label{eq:lml},
\end{equation}
where the vector ${\bm\Theta}\in\mathbb{R}^{\mathrm{{\rm N_D}}}$ contains all potential parameters that will be varied, whilst ${\bf D}({\bf X})\in\mathbb{R}^{\rm {\rm N_D}}$ is a \textit{descriptor vector} of the 
atomic coordinates ${\bf X}\in\mathbb{R}^{\rm3N}$. 
\change{The linear form 
(Eq. \ref{eq:lml}) encompasses linear sums of polynomial combinations of the descriptor functions, most commonly quadratic combinations such as those used in \texttt{qSNAP} \cite{Thompson_snap_2015} (Methods). Hyperparameters, 
namely the cutoff radius and choice of descriptor functions, are held constant throughout.}
Specific details on the descriptors employed are presented in Methods, though our refitting procedure is 
general to any LML implementation.\\

Our starting point for retraining is a potential parametrization $\bm\Theta_0$ that 
can \change{closely match} at least the DFT lattice and elastic constants;
we target an error of less than $1\%$. 
Whilst this is sufficient for elastically-matched hybrid simulations, 
in practice we start from a state-of-the-art parametrization \cite{goryaeva2021}, 
\change{the quadratically extended bispectrum implementation of the \texttt{MILADY} package \cite{milady},}
which is similar in functional form to \texttt{qSNAP} \cite{Thompson_snap_2015} 
\change{but has a modified fitting procedure, being strongly preconditioned by the linear-in-descriptors solution \cite{goryaeva2021}}. 
Use of an already optimized potential additionally
allows us to investigate the effect of our retraining procedure in more detail.

As discussed above and shown in figure \ref{fig:glide_summary}(c), using the original potential 
$\bm\Theta_0$ in QM/ML simulations will produce new ionic forces $\mathbf{F}^{\mathrm{QM/ML}}$ in 
the cluster region, with associated descriptor gradients ${\bm\nabla}\mathbf{D}^{\mathrm{QM/ML}}$. 
Our goal is to find a new parametrization ${\bm\Theta}={\bm\Theta}_0+\delta{\bm\Theta}$, 
which minimizes the error to $\mathbf{F}^{\mathrm{QM/ML}}$ whilst \textit{exactly} 
preserving `hard' properties such as elastic properties and approximately maintaining 
`soft' properties such as forces from high temperature MD that 
are important to avoid overfitting issues \cite{allen2021atomic}.\\

To define suitable constraints, we note that under weak, arbitrary homogeneous deformations, the 
total energy change of a lattice is uniquely determined by the elastic constants.
As illustrated in Figure \ref{fig:glide_summary}(e), we therefore simply 
subject a perfect lattice (here, bcc) to multiple shear/expansion deformations, 
collating the energy differences into a vector ${\bf t}_{\rm H}\in\mathbb{R}^{N_{\rm H}}$, with 
corresponding descriptor vectors collated into a rectangular
`design matrix' ${\bf A}_{\rm H}\in\mathbb{R}^{\rm N_H\times N_D}$. 
We are free to include additional properties alongside the elastic deformations-
for example, in the next section, we also include the structures for vacancy migration.
We provide a routine to generate these constraints using the Atomic Simulation Environment\cite{Hjorth_Larsen_2017} (Supplementary Material).
To exactly preserve the hard constraints we require the predicted values ${\bf t}_{\rm H}$ are unchanged, i.e.
\begin{equation}
    {\bf t}_{\rm H} \equiv {\bf A}_{\rm H}{\bm\Theta}_0 = 
    {\bf A}_{\rm H}\left[{\bm\Theta}_0+\delta{\bm\Theta}\right]
    ,\quad
    \Rightarrow
    {\bf A}_{\rm H}\delta{\bm\Theta} = {\bf0}.
    \label{eq:hard_constraints}
\end{equation}
A solution for ${\bf A}_{\rm H}\delta{\bm\Theta} = {\bf0}$ requires that 
the rank ${\rm R}_{\rm H}={\rm rank}({\bf A}_{\rm H})$ is less than the dimension 
${\rm N_D}$ of $\delta{\bm\Theta}$. This is a central motivation for using the 
quadratically extended descriptor formalism, which gives ${\rm N_D}\sim1500$ even 
when calculating less than a hundred descriptor functions per atom (Methods). For this 
descriptor choice, the elastic constraints alone gave ${\rm R}_{\rm H}=25$, 
rising to $37$ when including vacancy migration. To find a general solution, we apply
singular value decomposition (SVD) \cite{strang} to ${\bf A}_{\rm H}$, obtaining ${\rm R}_{\rm H}$ singular 
values, along with a set of ${\rm R}_{\rm H}$ orthonormal \change{right} singular vectors ${\bf v}_1,\dots,{\bf v}_{{\rm R}_{\rm H}}$. We note this procedure naturally eliminates any duplication of data when building ${\bf A}_{\rm H}$, meaning it is simple to build constraints for e.g. highly anisotropic elastic properties.  
Any right vector ${\bf w}\in\mathbb{R}^{\rm N_D}$ that has non-zero projection ${\bf A}_{\rm H}{\bf w} \neq {\bf0}$
can thus be expressed as a linear combination of the right singular vectors. 
Forming a projection matrix ${\bf P}_{\rm H} = \sum_m {\bf v}_m{\bf v}_m^\top
\in\mathbb{R}^{\rm N_D \times N_D}$,
A general solution can then be found by forming the \textit{null space} projection matrix
${\bf Q}_{\rm H}=\mathbb{I}-{\bf P}_{\rm H}$, such that ${\bf A}_{\rm H}{\bf Q}_{\rm H}{\bf v}={\bf 0}$ 
for \textit{any} vector ${\bf v}$. As a result, ${\delta\bm\Theta} = {\bf Q}_{\rm H}{\bf v}$ will 
always satisfy Eq. \ref{eq:hard_constraints}. 
\change{In principle, one can then search for the vector $\delta{\bm\Theta}$ that best matches QM/ML forces,
whilst also satisfying ${\bf P}_{\rm H}\delta{\bm\Theta}={\bf0}$. However, as ${\rm R}_{\rm H}\ll{\rm N_D}$, 
in practice, this regression procedure is vulnerable to overfitting. To correct for this, we additionally
require that the new refitted potential approximately 
preserves a set of properties of the original potential,
to provide further `soft' constraints. 
\ag{is it possible to indicate the order of error?}\tds{see below}\ag{OK}
We have found that including molecular dynamics trajectories 
of the bulk crystal at a range of temperatures (Figure \ref{fig:glide_summary}(d)) 
gives a `soft' targets ${\bf t}_{\rm S}$ and design
matrix ${\bf A}_{\rm S}$ gives ${\rm R}_{\rm S}\to{\rm N}_{\rm D}$ additional constraints, 
which we include as a term $\lambda_{\rm S}\|{\bf A}_{\rm S}{\bf Q}_{\rm H}\delta{\bm\Theta}\|^2$
alongside an additional ridge penalty, giving a final cost function}

\begin{align}
    \mathcal{L}({\delta\bm\Theta}) 
    &=
    \|\mathbf{F}^{\mathrm{QM/ML}}
    +
    {\bm\nabla}\mathbf{D}^{\mathrm{QM/ML}}
    \left[{\bm\Theta}_0+{\bf Q}_{\rm H}\delta{\bm\Theta}\right]
    \|^2
    \nonumber\\
    &+\lambda_{\rm S}\|{\bf A}_{\rm S}{\bf Q}_{\rm H}\delta{\bm\Theta}\|^2
    +
    \lambda_0\|\delta{\bm\Theta}\|^2,
\end{align}
where $\lambda_{\rm S}$ controls the weighting of soft constraints 
and $\lambda_0$ the standard ridge regularization. The minimum criterion 
$\delta\mathcal{L}=0$ becomes a linear equation ${\bf C}\delta{\bm\Theta}={\bf y}$, 
where the matrix ${\bf C}$ is always full-rank due to the presence of the ridge term,
permitting a direct solution. A Python implementation of this procedure is provided (Methods).
\tds{In fact, after rechecking the code it turns out we tried both (5+ months ago) but ended up just using a simple linear solve, as it gave slightly better control over the ridge parameter. Sorry for the confusion! I have kept the comments! Also, we have now seen that simply minimizing $\|{\bf A}_{\rm S}{\bf Q}_{\rm H}\delta\bm\Theta\|^2$ rather than $\|{\bf P}_{\rm S}{\bf Q}_{\rm H}\delta\bm\Theta\|^2$ can also give nice results, but we don't want to rerun the calculations... for the next paper!}
\change{Via grid search we chose $\lambda_{\rm S},\lambda_0$ to yield 
a mean error of less than 0.015 eV/$\rm\AA$ to the soft constraint forces and 
0.025 eV/$\rm\AA$ to the QM/ML forces, from an initial mean error of nearly 0.2 eV/$\rm\AA$. 
A detailed presentation of the refitting results are presented in the supplementary material.
We found this balance of hyperparameters controlled against overfitting whilst providing excellent 
reproduction of energy profiles (via force integration) and relaxed dislocation core structures. 
Further investigation of increasing potential complexity to further improve the retraining
will be the subject of future work. We now apply this method to study dislocations 
in our test material, tungsten.}

\subsection{Dislocation glide in bcc tungsten}

In bcc materials screw dislocations with Burgers vector $\bf{b} = \frac{1}{2}\langle$111$\rangle$ along closed packed $\langle$111$\rangle$ directions are the most ubiquitous. Movement of these dislocation is possible in few slip planes \cite{Weinberger2013} with \{110\} planes being dominant at low temperatures \cite{Clouet2021, Dezerald2016}. Formation of junction dislocations with Burgers vector along $\langle100\rangle$ direction occurs as a result reactions of type $\frac{1}{2}[1\bar{1}1] + \frac{1}{2}[11\bar{1}] = [100]$ during strain hardening \cite{Bulatov2006} or plasma exposure \cite{Guo2019}. Resulting $\langle100\rangle$ dislocation can glide in few planes including \{001\} and \{011\} depending on the geometry of the reaction \cite{Bertin2021}.

In this section we use an Embedded Atom Method (EAM) potential marked as ``EAM2'' from \cite{Marinica2013}, in addition to
a recent machine Linear Machine Learning (LML) potential from \cite{goryaeva2021} 
for our application material, tungsten. \change{The LML potential, whose general form is given in (\ref{eq:lml}), has state-of-the-art accuracy on a wide range of lattice, 
point defect and screw dislocation properties.}
These two potentials are employed to calculate the Peierls barriers for junction $\langle100\rangle$ dislocation in \{001\} and \{011\} planes. We analyse the performance of the potentials by comparing the results to QM/ML calculations using the LML potential for the embedding ML region, chosen as the LML has perfect matching of the QM elastic constants. \change{Total number of atoms in the cells were 4574 atoms for [100](00$\bar{1}$) model dislocation and 6396 for [100](011) dislocation. While QM/ML mapping consisted of 24 QM and 123 buffer atoms (147 atoms in DFT cluster) for [100](00$\bar{1}$) dislocation and 14 QM and 96 buffer atoms (110 atoms in DFT cluster) for [100](011) dislocation.}

Figures \ref{fig:glide_summary}(a,b) show Minimum Energy Paths (MEP) for Peierls barriers calculated with our modified force-only NEB routine (Methods).
For the \{001\} glide plane, figure \ref{fig:glide_summary}(a), the EAM and LML potentials differ significantly both in terms of amplitude and shape of the barrier while for \{011\} glide plane, figure \ref{fig:glide_summary}(b), the barriers are practically indistinguishable. 
Identical results were 
found using force integration and total energy differences for the NEB routine, \change{again confirming the accuracy of this approach \cite{Kermode_Swinburne_2017}.}
Both EAM and LML potentials predict the barrier for glide in \{001\} plane at least three times higher than in \{011\}. At the same time QM/ML results provide similar values for both glide planes around 0.4 eV. The glide barrier in \{001\} is overestimated by the potentials while the barrier in \{011\} is underestimated leading to qualitative disagreement with QM/ML reference data and overall poor performance of the potentials. It is important to note that this work QM/ML barriers are the only available QM reference data for this type of dislocations due to the large size of the simulation cell. \change{Previous work \cite{Kermode_Swinburne_2017,PhysRevMaterials.4.023601} has validated and performed extensive convergence checks for the virtual work NEB procedure detailed here.}

\change{
Access to the QM/ML forces allow us to produce a new LML parametrization using 
our constrained refitting procedure. 
The results are shown with green dots and dashed lines on Figures \ref{fig:glide_summary}(a,b).
The residual force errors of the retrained potential have very little influence on the resulting energy barriers, where an excellent agreement
can be seen. We emphasize that the refitting only targeted QM/ML forces, though the resulting LML energetic barriers can be 
calculated equivalently by force integration or total energy difference.} 
\change{
As the retrained LML potentials closely reproduce dislocation core properties and have essentially perfect matching of elastic constants, they can be used as a highly adaptable embedding medium. We performed QM/ML calculations with a spherical QM region, where the dislocation line crosses the QM/ML boundary (see figures
\ref{fig:vac_junction} and \ref{fig:imp_screw}(b)), the first time such calculations 
have been performed to our knowledge. We confirmed that a relaxing a long, 
straight dislocation in this manner, with a spherical QM region around a small section of the dislocation core, gave no appreciable change in structure, with {\it maximum} atomic displacement of less than 0.004 $\rm\AA$ between the ML and QM regions.}
In the following sections we exploit this refitting procedure to investigate vacancy segregation to prismatic dislocations, then Helium-induced core reconstruction of screw dislocations.

\subsection{Dislocation-vacancy segregation energy}
The biased diffusion of vacancies to dislocations is the primary source of non-conservative plastic deformation, a critically important process in creep deformation and post-irradiation annealing \cite{cottrell1967introduction}. Here we look at vacancy segregation at junction $\langle$100$\rangle$\{011\} dislocation as the case study for cross validation of 3D QM/ML coupling procedure against the original LML and a retrained parametrization. We retrained with hard 
constraints on elastic properties as described above and the vacancy migration pathway, 
both of which remain in essentially perfect agreement with reference DFT calculations. We confirmed that inclusion of the vacancy migration path in ${\bf A}_{\rm H}$ during the retraining left resulting core structure and Peierls barrier unchanged (Supp. Mat. \cite{sm}).

\begin{figure}
    \centering
    \includegraphics[width=0.5\textwidth]{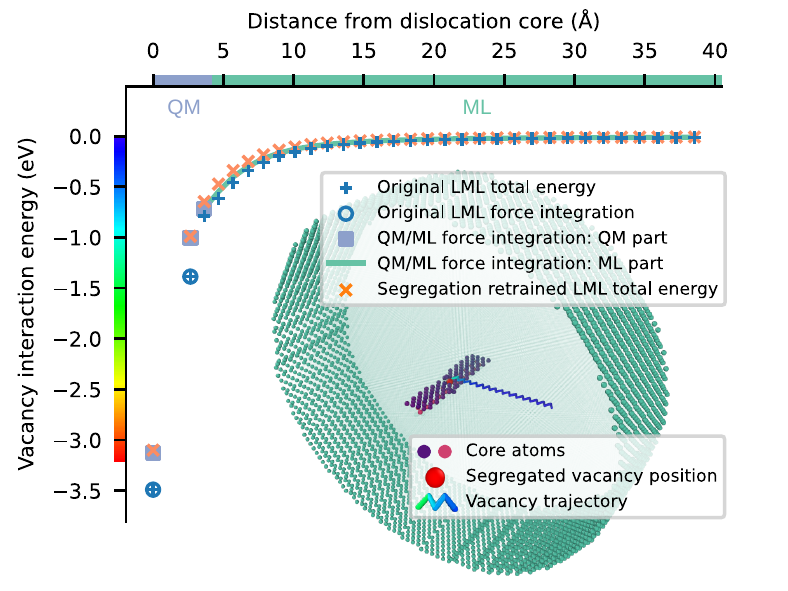}
    \caption{Energy profile for vacancy segregation to $[100](011)$ edge (also known as junction\cite{bertin2021core}) dislocations in tungsten calculated with multiple LML potentials and QM/ML force integration \cite{Kermode_Swinburne_2017,PhysRevMaterials.4.023601}. The full simulation contained over 70,000 atoms, with the vacancy migrating to a spherical QM region. The original LML potential\cite{goryaeva2021} results demonstrates the force integration procedure exactly agrees with total energy calculations. The QM/ML simulations (for which a total energy does not exist) employed a retrained LML potential from the NEB path shown in \ref{fig:glide_summary}, giving excellent reproduction of the core structure with hard constraints of elastic properties and the vacancy migration. We then further retrained this potential to additionally minimize the error to QM forces from the QM/ML segregation calculation. This final LML potential gives excellent prediction of the segregation energy.}
    \label{fig:vac_junction}
\end{figure}

In order to estimate vacancy segregation energy three configurations \change{containing 70,355 atoms each} with vacancy at increasing distances from the dislocation core were relaxed. After that eleven intermediate configurations between each pair of relaxed configurations were obtained by linear interpolation of atomic positions. A spherical QM region was kept the same for the all the configurations and consisted of the union of spheres centered on the relaxed vacancy positions \change{leading to 27 QM atoms and 351 Buffer atoms (378 atoms in final DFT cell)}. 
This large Buffer/QM ratio was chosen to ensure a highly converged solution; however, 
as we only employ a single k-point the total computational effort is much less than a 
comparably sized periodic calculation. 

The segregation energy was calculated via the virtual work principle, integrating forces along a composite path formed by linearly interpolating minima
separated by $\langle111\rangle$ jumps, as shown in figure \ref{fig:vac_junction}. Whilst it would be possible to perform NEB relaxations between these minima to additionally obtain migration barriers, this was omitted for computational expediency. 
The smoothness of the interaction energy across the QM/ML boundary confirms the exceptional matching of dislocation and elastic properties between retrained LML and QM region. 

The resulting value of segregation energy estimated by QM/ML approach is 3.1 eV, whilst the original LML potential gives 
3.5 eV, demonstrating the \change{reasonable} accuracy of this potential even for previously unseen structures. We calculated the segregation energy using both the total energy difference (blue crosses) and virtual work (blue hollow circles). The values for both techniques are identical, serving as a validation for energy difference estimation by virtual work principle.
We further explored the capabilities of the retraining procedure by extending the dislocation glide barrier targets with forces from spherical QM region for three QM/ML relaxed configurations.
The resulting segregation retrained potential is in excellent agreement to \change{the QM/ML reference result}. This demonstrates that the  QM/ML retraining procedure detailed here can not only provide 
advanced boundary matching but can also qualitatively expand the range of useful training structures for  
machine learning interatomic potentials, giving systematic improvement. 

\subsection{Impurity-induced core reconstruction of screw dislocations in tungsten}

\begin{figure*}
    \centering
    \includegraphics[width=0.9\textwidth]{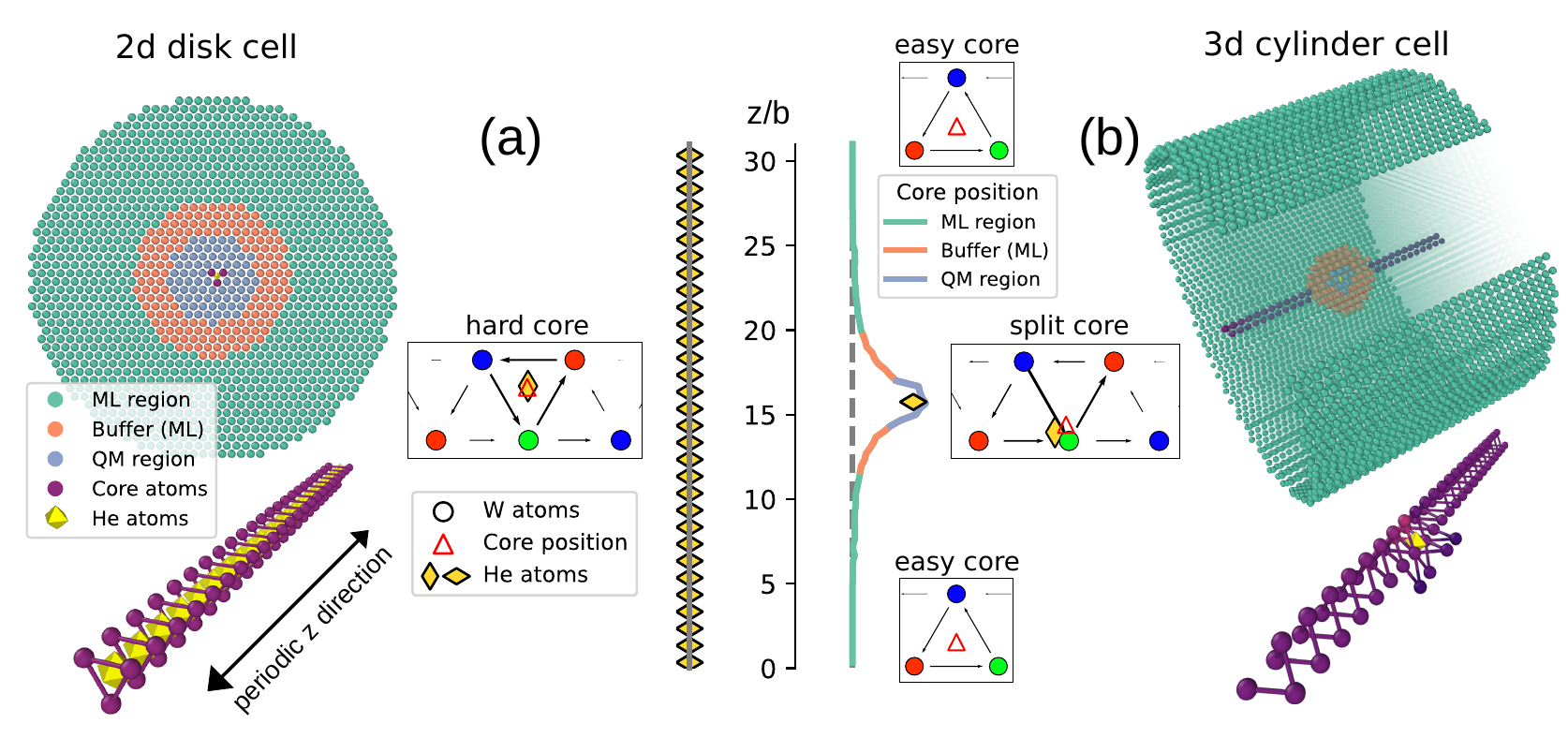}
    \caption{He-induced core reconstruction of screw dislocation cores obtained with two simulation geometries. Dislocation core structures are shown using 
    differential displacement maps\cite{Vitek1974}.
    (a) An atomic `disk' of primitive thickness (i.e. a single lattice plane) oriented perpendicular to the dislocation line with a circular QM region centered on the line. This geometry was used for the NEB calculations in figure \ref{fig:glide_summary}. Introduction of He stabilizes the `hard' core over the `easy' core, changing the dislocation migration trajectory\cite{Dezerald2016}.
    (b) A large cylindrical cell orientated along the line direction, with a spherical QM region centered on the He impurity. The He induces a qualitatively different reconstruction to the `split' core, which slowly relaxes back to the `easy' core far from the He center. We emphasise that for all the cases the `Buffer' region is shown purely for descriptive purposes; all atoms in this region are governed by the same LML potential as those in the ML region.}
    \label{fig:imp_screw}
\end{figure*}

In pure tungsten screw $\frac{1}{2}\langle$111$\rangle$ dislocations are characterised by so called `easy' core ~\cite{Vitek1974,Takeuchi1979,CAI_2004} shown by means of differential displacement map on the top and the bottom parts of figure \ref{fig:imp_screw}(b). The glide of screw dislocations is essentially the movement between two equivalent `easy' core configurations \cite{Ventelon2013}. Point defects segregated on a dislocation line can affect the relative stability of different core types thus changing the glide mechanism locally. In this section we consider the effect of single helium atom on the core stability of screw dislocation as a challenging application of 3D QM/ML coupling involving foreign atoms in the QM region. The original LML has an excellent agreement with DFT for the screw dislocation glide barrier and core structure \cite{goryaeva2021}.
The retraining procedure in this case only gave very small adjustments to the original parametrization, which 
primarily gave small changes in dislocation core structure,
with negligible changes in the Peierls barrier. Nevertheless, the retraining procedure was employed to ensure perfect matching for 3D spherical QM 
region. Importantly, in all cases the DFT region is sufficiently large that no tungsten atoms in the ML region interact directly with the He impurity, 
only indirectly through induced relaxations. 
\change{In the general case, where the direct interaction range is larger than the QM region, the ML 
potential would have to account for the impurity interaction at this range. }
Whilst we do not anticipate this to arise in most application settings, this will 
be further investigated in future work.

Figure \ref{fig:imp_screw}(a) shows the stabilisation of the `hard' core by He impurity atom while in pure material this type of core is unstable \cite{Dezerald2016}. These results are obtained using a computational cell containing a one $|\bf{b}|$ thick disk of atoms oriented perpendicular to the dislocation line. \change{The cell consists of 1,927 atoms with 79 QM atoms shown with blue spheres and 168 buffer atoms shown with orange spheres resulting in a DFT cell containing 247 atoms.} Periodic boundary conditions along z direction corresponding to the dislocation line yields in a model quasi infinite dislocation. However, when an impurity atom is added to this cell, it effectively models a dislocation fully decorated with He atoms shown at the bottom of figure \ref{fig:imp_screw}(a). 

Figure \ref{fig:imp_screw}(b) shows the results obtained with a cylinder cell \change{composed of 59,707 atoms} with a spherical QM region centered around the He atom. \change{The resulting DFT cluster contained 396 atoms with 24 atoms in QM region (blue atoms) and 373 atoms in Buffer (orange atoms).} The left part of the figure demonstrates the extracted dislocation core position by fitting theoretical displacement field to the displacement extracted from the relaxed atomic positions with $|\bf{b}|$/3 discretisation step \cite{Dezerald2016, HACHET2020481}. It can be seen that the in contrast to `disk' cell results He stabilises the `split' core locally. The dislocation goes back to 'easy' core configuration at distance of ~10 $|\bf{b}|$ from the impurity. \change{Similar relaxation lengths were obtained with fully periodic cells for carbon stabilised hard core in tungsten  \cite{HACHET2020481}. In this study cells up to 10 $|\bf{b}|$ length along dislocation line containing 1,350 atoms were considered. Extracted dislocation core position demonstrated that the dislocation does not completely recover back to easy core far from C atom even for  10 $|\bf{b}|$ cell.} This clearly demonstrates that it is essential to have a large cylinder configuration in order to capture effects of point defects on dislocations correctly. \change{It is important to note that carbon atom remained in the same position with increasing the length of the cell while the stabilised core position moved to the middle point on the straight path between hard and easy cores. We obtained significant difference of He atom position between `disk' and `cylinder' cells as well as the stabilised split core position largely deviates from the straight easy-hard core path.}
\pg{add comment about C and He migration barriers (1.47 vs 0.06 eV) here?} \mcm{naive point: also the interaction energies are not different between C and He ?  Is a question ... I do not know the literature ...  }\tds{Yes, for sure. The important point is the similarity of the spreading from line tension. We don't need to comment too much I think.}
The absence of jumps of the core position as the dislocation core exits the QM region confirms that we have achieved excellent matching of core properties between QM region and retrained LML potential resulting in a unique simulation tool.

\section{Conclusions}
In this paper, we have shown how state-of-the-art linear machine learning potentials can be 
used in advanced QM/ML simulations of extended lattice defects. As ML potentials can readily 
match the exact elastic properties of a reference DFT material, they are an ideal embedding 
medium for `traditional' open boundary calculations, which require only elastic matching 
at the QM boundary. We then used forces obtained from QM/ML simulations of dislocation glide
to retrain the potential, exploiting the linear form to geometrically ensure some 
subset properties, here elastic constants and vacancy migration energetics, remain constant. 
This allowed much more complex atomic configurations at the QM boundary, allowing 
for fully three dimensional simulations of impurity segregation to dislocations, where
long range reconstructions illustrated the importance of the method we present.  

We emphasize that the largest DFT simulations used in the present work contained 
less than four hundred atoms, with a single $k$-point due to the open boundaries.
This is approximately equivalent in computational effort to a 128 atom 
system with a sparse $3\times3\times3$ $k$-point grid, a routine 
calculation in modern theoretical materials science. The presented method 
is generally applicable for any extended defect, including grain boundaries 
or cracks, and offers many opportunities for the systematic improvement 
of machine learning potentials. 

\section{Acknowledgements}
PG and TDS gratefully recognize support from the Agence Nationale de Recherche, via the MeMoPas project ANR-19-CE46-0006-1 and the Centre Nationale de Recherce Scientifique, via a Jeunes Entrants grant of the Institut de Physique.
This work was granted access to the HPC resources of IDRIS under the allocation A0090910965 attributed by GENCI and the Computational Simulation Centre of the International Fusion Energy Research Centre in the Rokkasho Fusion Institute of QST (Aomori, Japan), under the Broader Approach grant AbInSeg. This work has been carried out within the framework of the EUROfusion consortium and has received funding from the Euratom research and training programme 2019-2020 under grant agreement No 633053. AMG, MCM and TDS acknowledge the support from TGCC-CCRT computer centres under the allocation ``Grand Challenge" no. 502 on Topaze.
JRK and PG acknowledge support from the UK Engineering and Physical Sciences Research Council (EPSRC) under grant numbers EP/R012474/1 and EP/R043612/1. Additional support was provided by the Leverhulme Trust under grant RPG-2017-191. We are grateful for computational support from the UK national high performance computing service, ARCHER, for which access was obtained via the UKCP consortium and funded by EPSRC grant reference EP/P022065/1. Additional computing facilities were provided by the Scientific Computing Research Technology Platform of the University of Warwick.

\section{Code Availability}
The force based QM/MM calculator is available as a part of the \texttt{ase.calculators.qmmm} module of the Atomic Simulation Environment (ASE) package \cite{Hjorth_Larsen_2017} as well as the used implementation of preconditioned minimisation algorithms are part of \texttt{ase.optimize.precon} module. The tools used to create and analyse atomistic dislocation configurations are available from the \texttt{matscipy.dislocation} module~\cite{matscipy}. 
A GitHub repository providing ASE-compatible routines to perform the retraining and reproduce all presented results is available at 
\url{https://github.com/marseille-matmol/LML-retrain}.

\section{Methods}

\subsection{Calculation of energies in QM/ML simulations}
    An important \change{aspect} of hybrid QM/ML simulations (and open boundary methods more generally) is 
    the need to have a `buffer' of sacrificial DFT atoms, suitably large to protect the target cluster from electronic free surface effects, determined through convergence tests \cite{Kermode_Swinburne_2017}. In this paper we use previously 
    validated parameters for tungsten \cite{PhysRevMaterials.4.023601}. 
    Here, the positions of buffer atoms are \change{inherited} 
    from atoms in the coupling medium, 
    as illustrated in figure \ref{fig:imp_screw}. One well known consequence of this procedure is that as 
    the electronic total energy cannot be partitioned between cluster and buffer, a total energy cannot be defined \cite{bernstein2009hybrid}. 
    However, the ionic (Hellman-Feynmann) forces can be unambiguously assigned, giving a total force vector $\mathbf{F}$
    for the system, which has been used to perform structural minimization and dynamics \change{for some time} \cite{bernstein2009hybrid}. In recent work \cite{Kermode_Swinburne_2017,PhysRevMaterials.4.023601}, we have partially lifted this limitation,  
    rigorously extracting energetic \textit{differences} between two configurations through the principle 
    of virtual work. This technique has been implemented in ASE as part of the NEB routine, to compute energies along minimum energy paths. The computational cost of NEB relaxation can be avoided if only end-to-end differences are desired, 
    as demonstrated below when calculating the dislocation-vacancy binding energy, following a linear interpolation
    between minima.
    For initial and final atomic configurations ${\bf X}_{\rm i,f}$, we construct some 
    smooth pathway ${\bf X}(\lambda)$, where $\lambda\in[0,1]$ is an affine parameter such that ${\bf X}(0)={\bf X}_{\rm i}$
    and ${\bf X}(1)={\bf X}_{\rm f}$, with corresponding ionic forces ${\bf F(\lambda')}$.
    The virtual work energy difference along the pathway then reads \cite{Kermode_Swinburne_2017,PhysRevMaterials.4.023601}
    \begin{equation}
        E(\lambda)-E(0) = \int_0^\lambda {\bf F}(\lambda')\frac{\rm d}{\rm d\lambda'}{\bf X}(\lambda'){\rm d}\lambda',
        \label{vw}
    \end{equation}
    where $({\rm d}/{\rm d\lambda'}){\bf X}(\lambda')$ is the pathway tangent.
    If performing a NEB relaxation, $E(\lambda)-E(0)$ will give the minimum energy profile. If the path is an unrelaxed
    interpolation between two minima, then only the total difference $E(1)-E(0)$ is typically of practical use. 
    Whilst not encountered in the present work, it is possible that a partial relaxation may be beneficial in some settings, 
    to avoid very large forces along the pathway that could cause quadrature issues in Eq. \ref{vw}.\\
    \change{
    The NEB calculations presented in figure \ref{fig:glide_summary} evaluated energy barriers from (\ref{vw}), interpolating the eleven intermediate images using a spline interpolation of forces and positions, as detailed in previous work \cite{Kermode_Swinburne_2017,PhysRevMaterials.4.023601}. The stopping force tolerance for NEB path optimisation using FIRE algorithm \cite{FIRE2006} was \(0.05\ \mathrm{eV/\AA} \).
    An implementation of our method is provided in the {\tt ASE}
    simulation package (Code Availability).
    Starting positions for the NEB relaxation were a obtained by linear interpolation between initial and final configurations 
    relaxed with preconditioned minimisation with adaptive step size selection \cite{Packwood2016,mones2018preconditioners,Makri2019} with a maximum force tolerance of \( 0.01\ \mathrm{eV/\AA} \).} 
    
    
\subsection{\change{VASP} DFT Parameters}
Density functional simulations were performed using \texttt{VASP}\change{ \cite{VASP}}.
The PBE generalised gradient approximation~\cite{PBE} was used to describe effects of electron exchange and correlation together with a projector augmented wave (PAW) basis set with a cut-off energy of \(550\) eV.
Occupancies were smeared with a Methfessel-Paxton scheme of order one with a \(0.1 \) eV smearing width.
The Brillouin zone was sampled with a \( 1 \times 1 \times 12 \) Monkhorst-Pack \textit{k}-point grid for the 2d cluster simulations periodic along the dislocation line and single \textit{k}-point was used for the calculations with 3d spherical QM regions.
The values of these parameters were chosen after a series of convergence tests on forces with a tolerance of
few meV/\text{\r{A}}.

\subsection{Machine learning potential}

\pg{We should probably say that for the relaxations we used \texttt{LAMMPS} and all the retrained potentials can be used ther as well. maybe in Code availability?}

We employ a quadratic extension of the bispectral descriptor as 
implemented in the \texttt{MILADY} potential package, first 
introduced as part of the \texttt{SNAP} family of LML potentials \cite{Thompson_snap_2015}.
The initial quadratic parametrization used a novel preconditioning procedure 
as presented in detail in a recent publication \cite{goryaeva2021}, 
to which we refer the reader for further information. 

Briefly, let ${\rm B}_{ji}({\bf X}),j\in[0,{\rm N_B}]$ be the ${\rm N_B}$
bispectral components for an atom $i$, along with a constant 
component ${\rm B}_{0i}\equiv1$. Only neighboring atoms 
within the cutoff distance (here 4.7$\rm\AA$) are included in the 
descriptor function calculation. 
The quadratically extended descriptor vector for the system then reads 
\begin{equation}
    {\bf D}({\bf X})
    =
    \sum_i
    {\bigoplus}_{j\geq k} {\rm B}_{ji}({\bf X}){\rm B}_{ki}({\bf X})
    \in\mathbb{R}^{1+2{\rm N_B}+{{\rm N_B}^2}/{2}}
    ,
    \label{eq:desc}
\end{equation}
where $\oplus$ indicates concatenation, giving $({\rm N_B}+1)({\rm N_B}+2)/2=1+2{\rm N_B}+{{\rm N_B}^2}/{2}$ components.
The quadratically extended form Eq. (\ref{eq:desc})
includes all terms linear in ${\rm B}$ (terms when $k=0$). The number of bispectrum components
${\rm N_B}$ is determined by an angular moment parameter 
$j_{max}=4$, giving ${\rm N_B}=55$. The original 
parametrization $\bm\Theta_0$ is determined in a two stage regression 
procedure
, named `quadratic noise' \cite{goryaeva2021}, where a fit is first
found using only the 55+1=56 linear combination of bispectrum components, 
which is then used to precondition a solution employing 
the above quadratic extension. 


\bibliography{bibliography}

\begin{thebibliography}{82}%
\makeatletter
\providecommand \@ifxundefined [1]{%
 \@ifx{#1\undefined}
}%
\providecommand \@ifnum [1]{%
 \ifnum #1\expandafter \@firstoftwo
 \else \expandafter \@secondoftwo
 \fi
}%
\providecommand \@ifx [1]{%
 \ifx #1\expandafter \@firstoftwo
 \else \expandafter \@secondoftwo
 \fi
}%
\providecommand \natexlab [1]{#1}%
\providecommand \enquote  [1]{``#1''}%
\providecommand \bibnamefont  [1]{#1}%
\providecommand \bibfnamefont [1]{#1}%
\providecommand \citenamefont [1]{#1}%
\providecommand \href@noop [0]{\@secondoftwo}%
\providecommand \href [0]{\begingroup \@sanitize@url \@href}%
\providecommand \@href[1]{\@@startlink{#1}\@@href}%
\providecommand \@@href[1]{\endgroup#1\@@endlink}%
\providecommand \@sanitize@url [0]{\catcode `\\12\catcode `\$12\catcode
  `\&12\catcode `\#12\catcode `\^12\catcode `\_12\catcode `\%12\relax}%
\providecommand \@@startlink[1]{}%
\providecommand \@@endlink[0]{}%
\providecommand \url  [0]{\begingroup\@sanitize@url \@url }%
\providecommand \@url [1]{\endgroup\@href {#1}{\urlprefix }}%
\providecommand \urlprefix  [0]{URL }%
\providecommand \Eprint [0]{\href }%
\providecommand \doibase [0]{http://dx.doi.org/}%
\providecommand \selectlanguage [0]{\@gobble}%
\providecommand \bibinfo  [0]{\@secondoftwo}%
\providecommand \bibfield  [0]{\@secondoftwo}%
\providecommand \translation [1]{[#1]}%
\providecommand \BibitemOpen [0]{}%
\providecommand \bibitemStop [0]{}%
\providecommand \bibitemNoStop [0]{.\EOS\space}%
\providecommand \EOS [0]{\spacefactor3000\relax}%
\providecommand \BibitemShut  [1]{\csname bibitem#1\endcsname}%
\let\auto@bib@innerbib\@empty
\bibitem [{\citenamefont {Hirth}\ and\ \citenamefont {Lothe}(1991)}]{Hirth}%
  \BibitemOpen
  \bibfield  {author} {\bibinfo {author} {\bibfnamefont {J.~P.}\ \bibnamefont
  {Hirth}}\ and\ \bibinfo {author} {\bibfnamefont {J.}~\bibnamefont {Lothe}},\
  }\href@noop {} {\emph {\bibinfo {title} {Theory Of Dislocations}}}\ (\bibinfo
   {publisher} {Malabar, FL Krieger},\ \bibinfo {year} {1991})\BibitemShut
  {NoStop}%
\bibitem [{\citenamefont {Leyson}\ \emph
  {et~al.}(2010{\natexlab{a}})\citenamefont {Leyson}, \citenamefont {Curtin},
  \citenamefont {Hector},\ and\ \citenamefont
  {Woodward}}]{leyson2010quantitative}%
  \BibitemOpen
  \bibfield  {author} {\bibinfo {author} {\bibfnamefont {G.~P.~M.}\
  \bibnamefont {Leyson}}, \bibinfo {author} {\bibfnamefont {W.~A.}\
  \bibnamefont {Curtin}}, \bibinfo {author} {\bibfnamefont {L.~G.}\
  \bibnamefont {Hector}}, \ and\ \bibinfo {author} {\bibfnamefont {C.~F.}\
  \bibnamefont {Woodward}},\ }\href@noop {} {\bibfield  {journal} {\bibinfo
  {journal} {Nature materials}\ }\textbf {\bibinfo {volume} {9}},\ \bibinfo
  {pages} {750} (\bibinfo {year} {2010}{\natexlab{a}})}\BibitemShut {NoStop}%
\bibitem [{\citenamefont {Rodney}\ \emph {et~al.}(2017)\citenamefont {Rodney},
  \citenamefont {Ventelon}, \citenamefont {Clouet}, \citenamefont
  {Pizzagalli},\ and\ \citenamefont {Willaime}}]{Rodney_2017}%
  \BibitemOpen
  \bibfield  {author} {\bibinfo {author} {\bibfnamefont {D.}~\bibnamefont
  {Rodney}}, \bibinfo {author} {\bibfnamefont {L.}~\bibnamefont {Ventelon}},
  \bibinfo {author} {\bibfnamefont {E.}~\bibnamefont {Clouet}}, \bibinfo
  {author} {\bibfnamefont {L.}~\bibnamefont {Pizzagalli}}, \ and\ \bibinfo
  {author} {\bibfnamefont {F.}~\bibnamefont {Willaime}},\ }\href {\doibase
  https://doi.org/10.1016/j.actamat.2016.09.049} {\bibfield  {journal}
  {\bibinfo  {journal} {Acta Materialia}\ }\textbf {\bibinfo {volume} {124}},\
  \bibinfo {pages} {633 } (\bibinfo {year} {2017})}\BibitemShut {NoStop}%
\bibitem [{\citenamefont {Hu}\ \emph {et~al.}(2017)\citenamefont {Hu},
  \citenamefont {Fellinger}, \citenamefont {Butler}, \citenamefont {Wang},
  \citenamefont {Darling}, \citenamefont {Kecskes}, \citenamefont {Trinkle},\
  and\ \citenamefont {Liu}}]{Hu2017}%
  \BibitemOpen
  \bibfield  {author} {\bibinfo {author} {\bibfnamefont {Y.-J.}\ \bibnamefont
  {Hu}}, \bibinfo {author} {\bibfnamefont {M.~R.}\ \bibnamefont {Fellinger}},
  \bibinfo {author} {\bibfnamefont {B.~G.}\ \bibnamefont {Butler}}, \bibinfo
  {author} {\bibfnamefont {Y.}~\bibnamefont {Wang}}, \bibinfo {author}
  {\bibfnamefont {K.~A.}\ \bibnamefont {Darling}}, \bibinfo {author}
  {\bibfnamefont {L.~J.}\ \bibnamefont {Kecskes}}, \bibinfo {author}
  {\bibfnamefont {D.~R.}\ \bibnamefont {Trinkle}}, \ and\ \bibinfo {author}
  {\bibfnamefont {Z.-K.}\ \bibnamefont {Liu}},\ }\href {\doibase
  https://doi.org/10.1016/j.actamat.2017.09.019} {\bibfield  {journal}
  {\bibinfo  {journal} {Acta Materialia}\ }\textbf {\bibinfo {volume} {141}},\
  \bibinfo {pages} {304} (\bibinfo {year} {2017})}\BibitemShut {NoStop}%
\bibitem [{\citenamefont {Schmidt}\ and\ \citenamefont
  {Avery}(1978)}]{schmidt1978complex}%
  \BibitemOpen
  \bibfield  {author} {\bibinfo {author} {\bibfnamefont {P.}~\bibnamefont
  {Schmidt}}\ and\ \bibinfo {author} {\bibfnamefont {D.~H.}\ \bibnamefont
  {Avery}},\ }\href@noop {} {\bibfield  {journal} {\bibinfo  {journal}
  {Science}\ }\textbf {\bibinfo {volume} {201}},\ \bibinfo {pages} {1085}
  (\bibinfo {year} {1978})}\BibitemShut {NoStop}%
\bibitem [{\citenamefont {Ventelon}\ \emph {et~al.}(2015)\citenamefont
  {Ventelon}, \citenamefont {L\"uthi}, \citenamefont {Clouet}, \citenamefont
  {Proville}, \citenamefont {Legrand}, \citenamefont {Rodney},\ and\
  \citenamefont {Willaime}}]{Ventelon_2015}%
  \BibitemOpen
  \bibfield  {author} {\bibinfo {author} {\bibfnamefont {L.}~\bibnamefont
  {Ventelon}}, \bibinfo {author} {\bibfnamefont {B.}~\bibnamefont {L\"uthi}},
  \bibinfo {author} {\bibfnamefont {E.}~\bibnamefont {Clouet}}, \bibinfo
  {author} {\bibfnamefont {L.}~\bibnamefont {Proville}}, \bibinfo {author}
  {\bibfnamefont {B.}~\bibnamefont {Legrand}}, \bibinfo {author} {\bibfnamefont
  {D.}~\bibnamefont {Rodney}}, \ and\ \bibinfo {author} {\bibfnamefont
  {F.}~\bibnamefont {Willaime}},\ }\href {\doibase 10.1103/PhysRevB.91.220102}
  {\bibfield  {journal} {\bibinfo  {journal} {Phys. Rev. B}\ }\textbf {\bibinfo
  {volume} {91}},\ \bibinfo {pages} {220102(R)} (\bibinfo {year}
  {2015})}\BibitemShut {NoStop}%
\bibitem [{\citenamefont {Cottrell}(1967)}]{cottrell1967introduction}%
  \BibitemOpen
  \bibfield  {author} {\bibinfo {author} {\bibfnamefont {A.}~\bibnamefont
  {Cottrell}},\ }\href@noop {} {\emph {\bibinfo {title} {An Introduction to
  Metallurgy}}}\ (\bibinfo  {publisher} {St. Martin's Press},\ \bibinfo {year}
  {1967})\BibitemShut {NoStop}%
\bibitem [{\citenamefont {Argon}(2008)}]{argon2008strengthening}%
  \BibitemOpen
  \bibfield  {author} {\bibinfo {author} {\bibfnamefont {A.}~\bibnamefont
  {Argon}},\ }\href@noop {} {\emph {\bibinfo {title} {Strengthening mechanisms
  in crystal plasticity}}},\ Vol.~\bibinfo {volume} {4}\ (\bibinfo  {publisher}
  {Oxford University Press on Demand},\ \bibinfo {year} {2008})\BibitemShut
  {NoStop}%
\bibitem [{\citenamefont {Yu}\ \emph {et~al.}(2015)\citenamefont {Yu},
  \citenamefont {Qi}, \citenamefont {Tsuru}, \citenamefont {Traylor},
  \citenamefont {Rugg}, \citenamefont {Morris}, \citenamefont {Asta},
  \citenamefont {Chrzan},\ and\ \citenamefont {Minor}}]{Yu2015}%
  \BibitemOpen
  \bibfield  {author} {\bibinfo {author} {\bibfnamefont {Q.}~\bibnamefont
  {Yu}}, \bibinfo {author} {\bibfnamefont {L.}~\bibnamefont {Qi}}, \bibinfo
  {author} {\bibfnamefont {T.}~\bibnamefont {Tsuru}}, \bibinfo {author}
  {\bibfnamefont {R.}~\bibnamefont {Traylor}}, \bibinfo {author} {\bibfnamefont
  {D.}~\bibnamefont {Rugg}}, \bibinfo {author} {\bibfnamefont {J.~W.}\
  \bibnamefont {Morris}}, \bibinfo {author} {\bibfnamefont {M.}~\bibnamefont
  {Asta}}, \bibinfo {author} {\bibfnamefont {D.~C.}\ \bibnamefont {Chrzan}}, \
  and\ \bibinfo {author} {\bibfnamefont {A.~M.}\ \bibnamefont {Minor}},\ }\href
  {\doibase 10.1126/science.1260485} {\bibfield  {journal} {\bibinfo  {journal}
  {Science}\ }\textbf {\bibinfo {volume} {347}} (\bibinfo {year} {2015}),\
  10.1126/science.1260485}\BibitemShut {NoStop}%
\bibitem [{\citenamefont {Leyson}\ \emph
  {et~al.}(2010{\natexlab{b}})\citenamefont {Leyson}, \citenamefont {Curtin},
  \citenamefont {Hector},\ and\ \citenamefont {Woodward}}]{Leyson2010}%
  \BibitemOpen
  \bibfield  {author} {\bibinfo {author} {\bibfnamefont {G.~P.~M.}\
  \bibnamefont {Leyson}}, \bibinfo {author} {\bibfnamefont {W.~A.}\
  \bibnamefont {Curtin}}, \bibinfo {author} {\bibfnamefont {L.~G.}\
  \bibnamefont {Hector}}, \ and\ \bibinfo {author} {\bibfnamefont {C.~F.}\
  \bibnamefont {Woodward}},\ }\href {\doibase 10.1038/nmat2813} {\bibfield
  {journal} {\bibinfo  {journal} {Nature Materials}\ }\textbf {\bibinfo
  {volume} {9}} (\bibinfo {year} {2010}{\natexlab{b}}),\
  10.1038/nmat2813}\BibitemShut {NoStop}%
\bibitem [{\citenamefont {Varvenne}\ \emph {et~al.}(2017)\citenamefont
  {Varvenne}, \citenamefont {Leyson}, \citenamefont {Ghazisaeidi},\ and\
  \citenamefont {Curtin}}]{Varvenne2017}%
  \BibitemOpen
  \bibfield  {author} {\bibinfo {author} {\bibfnamefont {C.}~\bibnamefont
  {Varvenne}}, \bibinfo {author} {\bibfnamefont {G.~P.}\ \bibnamefont
  {Leyson}}, \bibinfo {author} {\bibfnamefont {M.}~\bibnamefont {Ghazisaeidi}},
  \ and\ \bibinfo {author} {\bibfnamefont {W.~A.}\ \bibnamefont {Curtin}},\
  }\href {\doibase 10.1016/j.actamat.2016.09.046} {\bibfield  {journal}
  {\bibinfo  {journal} {Acta Materialia}\ }\textbf {\bibinfo {volume} {124}}
  (\bibinfo {year} {2017}),\ 10.1016/j.actamat.2016.09.046}\BibitemShut
  {NoStop}%
\bibitem [{\citenamefont {Nag}\ and\ \citenamefont {Curtin}(2020)}]{Nag2020}%
  \BibitemOpen
  \bibfield  {author} {\bibinfo {author} {\bibfnamefont {S.}~\bibnamefont
  {Nag}}\ and\ \bibinfo {author} {\bibfnamefont {W.~A.}\ \bibnamefont
  {Curtin}},\ }\href {\doibase 10.1016/J.ACTAMAT.2020.08.011} {\bibfield
  {journal} {\bibinfo  {journal} {Acta Materialia}\ }\textbf {\bibinfo {volume}
  {200}},\ \bibinfo {pages} {659} (\bibinfo {year} {2020})}\BibitemShut
  {NoStop}%
\bibitem [{\citenamefont {Zinkle}\ and\ \citenamefont
  {Was}(2013)}]{Zinkle2013}%
  \BibitemOpen
  \bibfield  {author} {\bibinfo {author} {\bibfnamefont {S.~J.}\ \bibnamefont
  {Zinkle}}\ and\ \bibinfo {author} {\bibfnamefont {G.~S.}\ \bibnamefont
  {Was}},\ }\href {\doibase 10.1016/j.actamat.2012.11.004} {\bibfield
  {journal} {\bibinfo  {journal} {Acta Materialia}\ }\textbf {\bibinfo {volume}
  {61}} (\bibinfo {year} {2013}),\ 10.1016/j.actamat.2012.11.004}\BibitemShut
  {NoStop}%
\bibitem [{\citenamefont {Zheng}\ \emph {et~al.}(2021)\citenamefont {Zheng},
  \citenamefont {Jian}, \citenamefont {Beyerlein},\ and\ \citenamefont
  {Han}}]{Zheng_VHe_hardening}%
  \BibitemOpen
  \bibfield  {author} {\bibinfo {author} {\bibfnamefont {R.-Y.}\ \bibnamefont
  {Zheng}}, \bibinfo {author} {\bibfnamefont {W.-R.}\ \bibnamefont {Jian}},
  \bibinfo {author} {\bibfnamefont {I.~J.}\ \bibnamefont {Beyerlein}}, \ and\
  \bibinfo {author} {\bibfnamefont {W.-Z.}\ \bibnamefont {Han}},\ }\href
  {\doibase 10.1021/acs.nanolett.1c01637} {\bibfield  {journal} {\bibinfo
  {journal} {Nano Letters}\ }\textbf {\bibinfo {volume} {21}},\ \bibinfo
  {pages} {5798} (\bibinfo {year} {2021})}\BibitemShut {NoStop}%
\bibitem [{\citenamefont {Li}\ \emph {et~al.}(2020)\citenamefont {Li},
  \citenamefont {Morgan}, \citenamefont {Terentyev}, \citenamefont {Ryelandt},
  \citenamefont {Favache}, \citenamefont {Wang}, \citenamefont {Wirtz},
  \citenamefont {Hoefnagels}, \citenamefont {Dommelen}, \citenamefont
  {Temmerman}, \citenamefont {Verbeken},\ and\ \citenamefont {Geers}}]{Li2020}%
  \BibitemOpen
  \bibfield  {author} {\bibinfo {author} {\bibfnamefont {Y.}~\bibnamefont
  {Li}}, \bibinfo {author} {\bibfnamefont {T.~W.}\ \bibnamefont {Morgan}},
  \bibinfo {author} {\bibfnamefont {D.}~\bibnamefont {Terentyev}}, \bibinfo
  {author} {\bibfnamefont {S.}~\bibnamefont {Ryelandt}}, \bibinfo {author}
  {\bibfnamefont {A.}~\bibnamefont {Favache}}, \bibinfo {author} {\bibfnamefont
  {S.}~\bibnamefont {Wang}}, \bibinfo {author} {\bibfnamefont {M.}~\bibnamefont
  {Wirtz}}, \bibinfo {author} {\bibfnamefont {J.~P.}\ \bibnamefont
  {Hoefnagels}}, \bibinfo {author} {\bibfnamefont {J.~A.~V.}\ \bibnamefont
  {Dommelen}}, \bibinfo {author} {\bibfnamefont {G.~D.}\ \bibnamefont
  {Temmerman}}, \bibinfo {author} {\bibfnamefont {K.}~\bibnamefont {Verbeken}},
  \ and\ \bibinfo {author} {\bibfnamefont {M.~G.}\ \bibnamefont {Geers}},\
  }\href {\doibase 10.1088/1741-4326/ab98a4} {\bibfield  {journal} {\bibinfo
  {journal} {Nuclear Fusion}\ }\textbf {\bibinfo {volume} {60}} (\bibinfo
  {year} {2020}),\ 10.1088/1741-4326/ab98a4}\BibitemShut {NoStop}%
\bibitem [{\citenamefont {Swinburne}\ and\ \citenamefont
  {Dudarev}(2018)}]{swinburne2018kink}%
  \BibitemOpen
  \bibfield  {author} {\bibinfo {author} {\bibfnamefont {T.~D.}\ \bibnamefont
  {Swinburne}}\ and\ \bibinfo {author} {\bibfnamefont {S.~L.}\ \bibnamefont
  {Dudarev}},\ }\href@noop {} {\bibfield  {journal} {\bibinfo  {journal}
  {Physical Review Materials}\ }\textbf {\bibinfo {volume} {2}},\ \bibinfo
  {pages} {073608} (\bibinfo {year} {2018})}\BibitemShut {NoStop}%
\bibitem [{\citenamefont {Ritchie}(2011)}]{Ritchie2011}%
  \BibitemOpen
  \bibfield  {author} {\bibinfo {author} {\bibfnamefont {R.~O.}\ \bibnamefont
  {Ritchie}},\ }\href {\doibase 10.1038/nmat3115} {\bibfield  {journal}
  {\bibinfo  {journal} {Nature Materials}\ }\textbf {\bibinfo {volume} {10}}
  (\bibinfo {year} {2011}),\ 10.1038/nmat3115}\BibitemShut {NoStop}%
\bibitem [{\citenamefont {Arakawa}\ \emph {et~al.}(2021)\citenamefont
  {Arakawa}, \citenamefont {Bergstrom}, \citenamefont {Caturla}, \citenamefont
  {Dudarev}, \citenamefont {Gao}, \citenamefont {Gilbert}, \citenamefont
  {Goryaeva}, \citenamefont {Hu}, \citenamefont {Hu}, \citenamefont {Kurtz}
  \emph {et~al.}}]{arakawa2021perspectives}%
  \BibitemOpen
  \bibfield  {author} {\bibinfo {author} {\bibfnamefont {K.}~\bibnamefont
  {Arakawa}}, \bibinfo {author} {\bibfnamefont {Z.}~\bibnamefont {Bergstrom}},
  \bibinfo {author} {\bibfnamefont {M.}~\bibnamefont {Caturla}}, \bibinfo
  {author} {\bibfnamefont {S.}~\bibnamefont {Dudarev}}, \bibinfo {author}
  {\bibfnamefont {F.}~\bibnamefont {Gao}}, \bibinfo {author} {\bibfnamefont
  {M.}~\bibnamefont {Gilbert}}, \bibinfo {author} {\bibfnamefont
  {A.}~\bibnamefont {Goryaeva}}, \bibinfo {author} {\bibfnamefont
  {S.}~\bibnamefont {Hu}}, \bibinfo {author} {\bibfnamefont {X.}~\bibnamefont
  {Hu}}, \bibinfo {author} {\bibfnamefont {R.~J.}\ \bibnamefont {Kurtz}},
  \emph {et~al.},\ }\href@noop {} {\bibfield  {journal} {\bibinfo  {journal}
  {Journal of Nuclear Materials}\ ,\ \bibinfo {pages} {153113}} (\bibinfo
  {year} {2021})}\BibitemShut {NoStop}%
\bibitem [{\citenamefont {Martin}(2004)}]{martin}%
  \BibitemOpen
  \bibfield  {author} {\bibinfo {author} {\bibfnamefont {R.~M.}\ \bibnamefont
  {Martin}},\ }\href@noop {} {\emph {\bibinfo {title} {{Electronic Structure:
  Basic Theory and Practical Methods}}}}\ (\bibinfo  {publisher} {Cambridge
  University Press},\ \bibinfo {year} {2004})\BibitemShut {NoStop}%
\bibitem [{\citenamefont {Woodward}(2005)}]{Woodward_2005}%
  \BibitemOpen
  \bibfield  {author} {\bibinfo {author} {\bibfnamefont {C.}~\bibnamefont
  {Woodward}},\ }\href {\doibase https://doi.org/10.1016/j.msea.2005.03.039}
  {\bibfield  {journal} {\bibinfo  {journal} {Materials Science and
  Engineering: A}\ }\textbf {\bibinfo {volume} {400-401}},\ \bibinfo {pages}
  {59 } (\bibinfo {year} {2005})},\ \bibinfo {note} {dislocations
  2004}\BibitemShut {NoStop}%
\bibitem [{\citenamefont {Ventelon}\ and\ \citenamefont
  {Willaime}(2007)}]{Ventelon2007}%
  \BibitemOpen
  \bibfield  {author} {\bibinfo {author} {\bibfnamefont {L.}~\bibnamefont
  {Ventelon}}\ and\ \bibinfo {author} {\bibfnamefont {F.}~\bibnamefont
  {Willaime}},\ }\href {\doibase 10.1007/s10820-007-9064-y} {\bibfield
  {journal} {\bibinfo  {journal} {Journal of Computer-Aided Materials Design}\
  }\textbf {\bibinfo {volume} {14}},\ \bibinfo {pages} {85} (\bibinfo {year}
  {2007})}\BibitemShut {NoStop}%
\bibitem [{\citenamefont {Dezerald}\ \emph {et~al.}(2015)\citenamefont
  {Dezerald}, \citenamefont {Proville}, \citenamefont {Ventelon}, \citenamefont
  {Willaime},\ and\ \citenamefont {Rodney}}]{dezerald2015first}%
  \BibitemOpen
  \bibfield  {author} {\bibinfo {author} {\bibfnamefont {L.}~\bibnamefont
  {Dezerald}}, \bibinfo {author} {\bibfnamefont {L.}~\bibnamefont {Proville}},
  \bibinfo {author} {\bibfnamefont {L.}~\bibnamefont {Ventelon}}, \bibinfo
  {author} {\bibfnamefont {F.}~\bibnamefont {Willaime}}, \ and\ \bibinfo
  {author} {\bibfnamefont {D.}~\bibnamefont {Rodney}},\ }\href@noop {}
  {\bibfield  {journal} {\bibinfo  {journal} {Physical Review B}\ }\textbf
  {\bibinfo {volume} {91}},\ \bibinfo {pages} {094105} (\bibinfo {year}
  {2015})}\BibitemShut {NoStop}%
\bibitem [{\citenamefont {Dezerald}\ \emph {et~al.}(2016)\citenamefont
  {Dezerald}, \citenamefont {Rodney}, \citenamefont {Clouet}, \citenamefont
  {Ventelon},\ and\ \citenamefont {Willaime}}]{Dezerald2016}%
  \BibitemOpen
  \bibfield  {author} {\bibinfo {author} {\bibfnamefont {L.}~\bibnamefont
  {Dezerald}}, \bibinfo {author} {\bibfnamefont {D.}~\bibnamefont {Rodney}},
  \bibinfo {author} {\bibfnamefont {E.}~\bibnamefont {Clouet}}, \bibinfo
  {author} {\bibfnamefont {L.}~\bibnamefont {Ventelon}}, \ and\ \bibinfo
  {author} {\bibfnamefont {F.}~\bibnamefont {Willaime}},\ }\href
  {https://doi.org/10.1038/ncomms11695} {\bibfield  {journal} {\bibinfo
  {journal} {Nature Communications}\ }\textbf {\bibinfo {volume} {7}},\
  \bibinfo {pages} {11695 EP } (\bibinfo {year} {2016})}\BibitemShut {NoStop}%
\bibitem [{\citenamefont {Clouet}\ \emph {et~al.}(2015)\citenamefont {Clouet},
  \citenamefont {Caillard}, \citenamefont {Chaari}, \citenamefont {Onimus},\
  and\ \citenamefont {Rodney}}]{clouet2015dislocation}%
  \BibitemOpen
  \bibfield  {author} {\bibinfo {author} {\bibfnamefont {E.}~\bibnamefont
  {Clouet}}, \bibinfo {author} {\bibfnamefont {D.}~\bibnamefont {Caillard}},
  \bibinfo {author} {\bibfnamefont {N.}~\bibnamefont {Chaari}}, \bibinfo
  {author} {\bibfnamefont {F.}~\bibnamefont {Onimus}}, \ and\ \bibinfo {author}
  {\bibfnamefont {D.}~\bibnamefont {Rodney}},\ }\href@noop {} {\bibfield
  {journal} {\bibinfo  {journal} {Nature materials}\ }\textbf {\bibinfo
  {volume} {14}},\ \bibinfo {pages} {931} (\bibinfo {year} {2015})}\BibitemShut
  {NoStop}%
\bibitem [{\citenamefont {Hachet}\ \emph {et~al.}(2020)\citenamefont {Hachet},
  \citenamefont {Ventelon}, \citenamefont {Willaime},\ and\ \citenamefont
  {Clouet}}]{HACHET2020481}%
  \BibitemOpen
  \bibfield  {author} {\bibinfo {author} {\bibfnamefont {G.}~\bibnamefont
  {Hachet}}, \bibinfo {author} {\bibfnamefont {L.}~\bibnamefont {Ventelon}},
  \bibinfo {author} {\bibfnamefont {F.}~\bibnamefont {Willaime}}, \ and\
  \bibinfo {author} {\bibfnamefont {E.}~\bibnamefont {Clouet}},\ }\href
  {\doibase https://doi.org/10.1016/j.actamat.2020.09.014} {\bibfield
  {journal} {\bibinfo  {journal} {Acta Materialia}\ }\textbf {\bibinfo {volume}
  {200}},\ \bibinfo {pages} {481} (\bibinfo {year} {2020})}\BibitemShut
  {NoStop}%
\bibitem [{\citenamefont {Woodward}\ and\ \citenamefont
  {Rao}(2002)}]{Woodward_2002}%
  \BibitemOpen
  \bibfield  {author} {\bibinfo {author} {\bibfnamefont {C.}~\bibnamefont
  {Woodward}}\ and\ \bibinfo {author} {\bibfnamefont {S.~I.}\ \bibnamefont
  {Rao}},\ }\href {\doibase 10.1103/PhysRevLett.88.216402} {\bibfield
  {journal} {\bibinfo  {journal} {Phys. Rev. Lett.}\ }\textbf {\bibinfo
  {volume} {88}},\ \bibinfo {pages} {216402} (\bibinfo {year}
  {2002})}\BibitemShut {NoStop}%
\bibitem [{\citenamefont {Kermode}\ \emph {et~al.}(2008)\citenamefont
  {Kermode}, \citenamefont {Albaret}, \citenamefont {Sherman}, \citenamefont
  {Bernstein}, \citenamefont {Gumbsch}, \citenamefont {Payne}, \citenamefont
  {Cs{\'a}nyi},\ and\ \citenamefont {De~Vita}}]{Kermode2008}%
  \BibitemOpen
  \bibfield  {author} {\bibinfo {author} {\bibfnamefont {J.~R.}\ \bibnamefont
  {Kermode}}, \bibinfo {author} {\bibfnamefont {T.}~\bibnamefont {Albaret}},
  \bibinfo {author} {\bibfnamefont {D.}~\bibnamefont {Sherman}}, \bibinfo
  {author} {\bibfnamefont {N.}~\bibnamefont {Bernstein}}, \bibinfo {author}
  {\bibfnamefont {P.}~\bibnamefont {Gumbsch}}, \bibinfo {author} {\bibfnamefont
  {M.~C.}\ \bibnamefont {Payne}}, \bibinfo {author} {\bibfnamefont
  {G.}~\bibnamefont {Cs{\'a}nyi}}, \ and\ \bibinfo {author} {\bibfnamefont
  {A.}~\bibnamefont {De~Vita}},\ }\href {\doibase 10.1038/nature07297}
  {\bibfield  {journal} {\bibinfo  {journal} {Nature}\ }\textbf {\bibinfo
  {volume} {455}},\ \bibinfo {pages} {1224} (\bibinfo {year}
  {2008})}\BibitemShut {NoStop}%
\bibitem [{\citenamefont {Fellinger}\ \emph {et~al.}(2018)\citenamefont
  {Fellinger}, \citenamefont {Tan}, \citenamefont {Hector},\ and\ \citenamefont
  {Trinkle}}]{Trinkle_2018}%
  \BibitemOpen
  \bibfield  {author} {\bibinfo {author} {\bibfnamefont {M.~R.}\ \bibnamefont
  {Fellinger}}, \bibinfo {author} {\bibfnamefont {A.~M.~Z.}\ \bibnamefont
  {Tan}}, \bibinfo {author} {\bibfnamefont {L.~G.}\ \bibnamefont {Hector}}, \
  and\ \bibinfo {author} {\bibfnamefont {D.~R.}\ \bibnamefont {Trinkle}},\
  }\href {\doibase 10.1103/PhysRevMaterials.2.113605} {\bibfield  {journal}
  {\bibinfo  {journal} {Phys. Rev. Materials}\ }\textbf {\bibinfo {volume}
  {2}},\ \bibinfo {pages} {113605} (\bibinfo {year} {2018})}\BibitemShut
  {NoStop}%
\bibitem [{\citenamefont {Swinburne}\ and\ \citenamefont
  {Kermode}(2017)}]{Kermode_Swinburne_2017}%
  \BibitemOpen
  \bibfield  {author} {\bibinfo {author} {\bibfnamefont {T.~D.}\ \bibnamefont
  {Swinburne}}\ and\ \bibinfo {author} {\bibfnamefont {J.~R.}\ \bibnamefont
  {Kermode}},\ }\href {\doibase 10.1103/PhysRevB.96.144102} {\bibfield
  {journal} {\bibinfo  {journal} {Phys. Rev. B}\ }\textbf {\bibinfo {volume}
  {96}},\ \bibinfo {pages} {144102} (\bibinfo {year} {2017})}\BibitemShut
  {NoStop}%
\bibitem [{\citenamefont {Bernstein}\ \emph {et~al.}(2009)\citenamefont
  {Bernstein}, \citenamefont {Kermode},\ and\ \citenamefont
  {Csanyi}}]{bernstein2009hybrid}%
  \BibitemOpen
  \bibfield  {author} {\bibinfo {author} {\bibfnamefont {N.}~\bibnamefont
  {Bernstein}}, \bibinfo {author} {\bibfnamefont {J.~R.}\ \bibnamefont
  {Kermode}}, \ and\ \bibinfo {author} {\bibfnamefont {G.}~\bibnamefont
  {Csanyi}},\ }\href@noop {} {\bibfield  {journal} {\bibinfo  {journal}
  {Reports on Progress in Physics}\ }\textbf {\bibinfo {volume} {72}},\
  \bibinfo {pages} {026501} (\bibinfo {year} {2009})}\BibitemShut {NoStop}%
\bibitem [{\citenamefont {Grigorev}\ \emph {et~al.}(2020)\citenamefont
  {Grigorev}, \citenamefont {Swinburne},\ and\ \citenamefont
  {Kermode}}]{PhysRevMaterials.4.023601}%
  \BibitemOpen
  \bibfield  {author} {\bibinfo {author} {\bibfnamefont {P.}~\bibnamefont
  {Grigorev}}, \bibinfo {author} {\bibfnamefont {T.~D.}\ \bibnamefont
  {Swinburne}}, \ and\ \bibinfo {author} {\bibfnamefont {J.~R.}\ \bibnamefont
  {Kermode}},\ }\href {\doibase 10.1103/PhysRevMaterials.4.023601} {\bibfield
  {journal} {\bibinfo  {journal} {Phys. Rev. Materials}\ }\textbf {\bibinfo
  {volume} {4}},\ \bibinfo {pages} {023601} (\bibinfo {year}
  {2020})}\BibitemShut {NoStop}%
\bibitem [{\citenamefont {Shao}(1993)}]{shao1993linear}%
  \BibitemOpen
  \bibfield  {author} {\bibinfo {author} {\bibfnamefont {J.}~\bibnamefont
  {Shao}},\ }\href@noop {} {\bibfield  {journal} {\bibinfo  {journal} {Journal
  of the American statistical Association}\ }\textbf {\bibinfo {volume} {88}},\
  \bibinfo {pages} {486} (\bibinfo {year} {1993})}\BibitemShut {NoStop}%
\bibitem [{\citenamefont {Srivastava}\ \emph {et~al.}(2014)\citenamefont
  {Srivastava}, \citenamefont {Hinton}, \citenamefont {Krizhevsky},
  \citenamefont {Sutskever},\ and\ \citenamefont
  {Salakhutdinov}}]{srivastava2014dropout}%
  \BibitemOpen
  \bibfield  {author} {\bibinfo {author} {\bibfnamefont {N.}~\bibnamefont
  {Srivastava}}, \bibinfo {author} {\bibfnamefont {G.}~\bibnamefont {Hinton}},
  \bibinfo {author} {\bibfnamefont {A.}~\bibnamefont {Krizhevsky}}, \bibinfo
  {author} {\bibfnamefont {I.}~\bibnamefont {Sutskever}}, \ and\ \bibinfo
  {author} {\bibfnamefont {R.}~\bibnamefont {Salakhutdinov}},\ }\href@noop {}
  {\bibfield  {journal} {\bibinfo  {journal} {The journal of machine learning
  research}\ }\textbf {\bibinfo {volume} {15}},\ \bibinfo {pages} {1929}
  (\bibinfo {year} {2014})}\BibitemShut {NoStop}%
\bibitem [{\citenamefont {MacKay}(1992)}]{mackay1992bayesian}%
  \BibitemOpen
  \bibfield  {author} {\bibinfo {author} {\bibfnamefont {D.~J.}\ \bibnamefont
  {MacKay}},\ }\href@noop {} {\bibfield  {journal} {\bibinfo  {journal} {Neural
  computation}\ }\textbf {\bibinfo {volume} {4}},\ \bibinfo {pages} {415}
  (\bibinfo {year} {1992})}\BibitemShut {NoStop}%
\bibitem [{\citenamefont {Goryaeva}\ \emph {et~al.}(2019)\citenamefont
  {Goryaeva}, \citenamefont {Maillet},\ and\ \citenamefont
  {Marinica}}]{goryaeva2019towards}%
  \BibitemOpen
  \bibfield  {author} {\bibinfo {author} {\bibfnamefont {A.~M.}\ \bibnamefont
  {Goryaeva}}, \bibinfo {author} {\bibfnamefont {J.-B.}\ \bibnamefont
  {Maillet}}, \ and\ \bibinfo {author} {\bibfnamefont {M.-C.}\ \bibnamefont
  {Marinica}},\ }\href@noop {} {\bibfield  {journal} {\bibinfo  {journal}
  {Computational Materials Science}\ }\textbf {\bibinfo {volume} {166}},\
  \bibinfo {pages} {200} (\bibinfo {year} {2019})}\BibitemShut {NoStop}%
\bibitem [{\citenamefont {Deringer}\ \emph {et~al.}(2021)\citenamefont
  {Deringer}, \citenamefont {Bart{\'o}k}, \citenamefont {Bernstein},
  \citenamefont {Wilkins}, \citenamefont {Ceriotti},\ and\ \citenamefont
  {Cs{\'a}nyi}}]{deringer2021gaussian}%
  \BibitemOpen
  \bibfield  {author} {\bibinfo {author} {\bibfnamefont {V.~L.}\ \bibnamefont
  {Deringer}}, \bibinfo {author} {\bibfnamefont {A.~P.}\ \bibnamefont
  {Bart{\'o}k}}, \bibinfo {author} {\bibfnamefont {N.}~\bibnamefont
  {Bernstein}}, \bibinfo {author} {\bibfnamefont {D.~M.}\ \bibnamefont
  {Wilkins}}, \bibinfo {author} {\bibfnamefont {M.}~\bibnamefont {Ceriotti}}, \
  and\ \bibinfo {author} {\bibfnamefont {G.}~\bibnamefont {Cs{\'a}nyi}},\
  }\href@noop {} {\bibfield  {journal} {\bibinfo  {journal} {Chemical Reviews}\
  }\textbf {\bibinfo {volume} {121}},\ \bibinfo {pages} {10073} (\bibinfo
  {year} {2021})}\BibitemShut {NoStop}%
\bibitem [{\citenamefont {Mishin}(2021)}]{mishin2021machine}%
  \BibitemOpen
  \bibfield  {author} {\bibinfo {author} {\bibfnamefont {Y.}~\bibnamefont
  {Mishin}},\ }\href@noop {} {\bibfield  {journal} {\bibinfo  {journal} {Acta
  Materialia}\ }\textbf {\bibinfo {volume} {214}},\ \bibinfo {pages} {116980}
  (\bibinfo {year} {2021})}\BibitemShut {NoStop}%
\bibitem [{\citenamefont {Onat}\ \emph {et~al.}(2020)\citenamefont {Onat},
  \citenamefont {Ortner},\ and\ \citenamefont {Kermode}}]{onat2020sensitivity}%
  \BibitemOpen
  \bibfield  {author} {\bibinfo {author} {\bibfnamefont {B.}~\bibnamefont
  {Onat}}, \bibinfo {author} {\bibfnamefont {C.}~\bibnamefont {Ortner}}, \ and\
  \bibinfo {author} {\bibfnamefont {J.~R.}\ \bibnamefont {Kermode}},\
  }\href@noop {} {\bibfield  {journal} {\bibinfo  {journal} {The Journal of
  Chemical Physics}\ }\textbf {\bibinfo {volume} {153}},\ \bibinfo {pages}
  {144106} (\bibinfo {year} {2020})}\BibitemShut {NoStop}%
\bibitem [{\citenamefont {Unke}\ \emph {et~al.}(2021)\citenamefont {Unke},
  \citenamefont {Chmiela}, \citenamefont {Sauceda}, \citenamefont {Gastegger},
  \citenamefont {Poltavsky}, \citenamefont {Sch\"utt}, \citenamefont
  {Tkatchenko},\ and\ \citenamefont {M\"uller}}]{Unke_machine_2021}%
  \BibitemOpen
  \bibfield  {author} {\bibinfo {author} {\bibfnamefont {O.~T.}\ \bibnamefont
  {Unke}}, \bibinfo {author} {\bibfnamefont {S.}~\bibnamefont {Chmiela}},
  \bibinfo {author} {\bibfnamefont {H.~E.}\ \bibnamefont {Sauceda}}, \bibinfo
  {author} {\bibfnamefont {M.}~\bibnamefont {Gastegger}}, \bibinfo {author}
  {\bibfnamefont {I.}~\bibnamefont {Poltavsky}}, \bibinfo {author}
  {\bibfnamefont {K.~T.}\ \bibnamefont {Sch\"utt}}, \bibinfo {author}
  {\bibfnamefont {A.}~\bibnamefont {Tkatchenko}}, \ and\ \bibinfo {author}
  {\bibfnamefont {K.-R.}\ \bibnamefont {M\"uller}},\ }\href {\doibase
  10.1021/acs.chemrev.0c01111} {\bibfield  {journal} {\bibinfo  {journal}
  {Chem. Rev.}\ }\textbf {\bibinfo {volume} {121}},\ \bibinfo {pages} {10142 }
  (\bibinfo {year} {2021})}\BibitemShut {NoStop}%
\bibitem [{\citenamefont {Bart\'ok}(2009)}]{bartok_thesis}%
  \BibitemOpen
  \bibfield  {author} {\bibinfo {author} {\bibfnamefont {A.~P.}\ \bibnamefont
  {Bart\'ok}},\ }\emph {\bibinfo {title} {Gaussian Approximation Potential : an
  interatomic potential derived from first principles Quantum Mechanics}},\
  \href@noop {} {Ph.D. thesis},\ \bibinfo  {school} {University of Cambridge}
  (\bibinfo {year} {2009})\BibitemShut {NoStop}%
\bibitem [{\citenamefont {Bart\'ok}\ \emph {et~al.}(2010)\citenamefont
  {Bart\'ok}, \citenamefont {Payne}, \citenamefont {Kondor},\ and\
  \citenamefont {Cs\'anyi}}]{bartok2010}%
  \BibitemOpen
  \bibfield  {author} {\bibinfo {author} {\bibfnamefont {A.~P.}\ \bibnamefont
  {Bart\'ok}}, \bibinfo {author} {\bibfnamefont {M.~C.}\ \bibnamefont {Payne}},
  \bibinfo {author} {\bibfnamefont {R.}~\bibnamefont {Kondor}}, \ and\ \bibinfo
  {author} {\bibfnamefont {G.}~\bibnamefont {Cs\'anyi}},\ }\href {\doibase
  10.1103/PhysRevLett.104.136403} {\bibfield  {journal} {\bibinfo  {journal}
  {Phys. Rev. Lett.}\ }\textbf {\bibinfo {volume} {104}},\ \bibinfo {pages}
  {136403} (\bibinfo {year} {2010})}\BibitemShut {NoStop}%
\bibitem [{\citenamefont {Behler}\ and\ \citenamefont
  {Parrinello}(2007)}]{behler2007}%
  \BibitemOpen
  \bibfield  {author} {\bibinfo {author} {\bibfnamefont {J.}~\bibnamefont
  {Behler}}\ and\ \bibinfo {author} {\bibfnamefont {M.}~\bibnamefont
  {Parrinello}},\ }\href@noop {} {\bibfield  {journal} {\bibinfo  {journal}
  {Physical review letters}\ }\textbf {\bibinfo {volume} {98}},\ \bibinfo
  {pages} {146401} (\bibinfo {year} {2007})}\BibitemShut {NoStop}%
\bibitem [{\citenamefont {Thompson}\ \emph {et~al.}(2015)\citenamefont
  {Thompson}, \citenamefont {Swiler}, \citenamefont {Trott}, \citenamefont
  {Foiles},\ and\ \citenamefont {Tucker}}]{Thompson_snap_2015}%
  \BibitemOpen
  \bibfield  {author} {\bibinfo {author} {\bibfnamefont {A.~P.}\ \bibnamefont
  {Thompson}}, \bibinfo {author} {\bibfnamefont {L.~P.}\ \bibnamefont
  {Swiler}}, \bibinfo {author} {\bibfnamefont {C.~R.}\ \bibnamefont {Trott}},
  \bibinfo {author} {\bibfnamefont {S.~M.}\ \bibnamefont {Foiles}}, \ and\
  \bibinfo {author} {\bibfnamefont {G.~J.}\ \bibnamefont {Tucker}},\ }\href
  {\doibase 10.1016/j.jcp.2014.12.018} {\bibfield  {journal} {\bibinfo
  {journal} {J. Comp. Phys.}\ }\textbf {\bibinfo {volume} {285}},\ \bibinfo
  {pages} {316} (\bibinfo {year} {2015})}\BibitemShut {NoStop}%
\bibitem [{\citenamefont {Shapeev}(2016)}]{Shapeev_MTP}%
  \BibitemOpen
  \bibfield  {author} {\bibinfo {author} {\bibfnamefont {A.}~\bibnamefont
  {Shapeev}},\ }\href {\doibase 10.1137/15M1054183} {\bibfield  {journal}
  {\bibinfo  {journal} {Multiscale Model. Sim.}\ }\textbf {\bibinfo {volume}
  {14}},\ \bibinfo {pages} {1153} (\bibinfo {year} {2016})}\BibitemShut
  {NoStop}%
\bibitem [{\citenamefont {Podryabinkin}\ and\ \citenamefont
  {Shapeev}(2017)}]{Shapeev_MTP2017}%
  \BibitemOpen
  \bibfield  {author} {\bibinfo {author} {\bibfnamefont {E.~V.}\ \bibnamefont
  {Podryabinkin}}\ and\ \bibinfo {author} {\bibfnamefont {A.~V.}\ \bibnamefont
  {Shapeev}},\ }\href {\doibase 10.1016/j.commatsci.2017.08.031} {\bibfield
  {journal} {\bibinfo  {journal} {Comput. Mater. Sci.}\ }\textbf {\bibinfo
  {volume} {140}},\ \bibinfo {pages} {171} (\bibinfo {year}
  {2017})}\BibitemShut {NoStop}%
\bibitem [{\citenamefont {Goryaeva}\ \emph
  {et~al.}(2021{\natexlab{a}})\citenamefont {Goryaeva}, \citenamefont
  {D\'er\`es}, \citenamefont {Lapointe}, \citenamefont {Grigorev},
  \citenamefont {Swinburne}, \citenamefont {Kermode}, \citenamefont {Ventelon},
  \citenamefont {Baima},\ and\ \citenamefont {Marinica}}]{goryaeva2021}%
  \BibitemOpen
  \bibfield  {author} {\bibinfo {author} {\bibfnamefont {A.~M.}\ \bibnamefont
  {Goryaeva}}, \bibinfo {author} {\bibfnamefont {J.}~\bibnamefont {D\'er\`es}},
  \bibinfo {author} {\bibfnamefont {C.}~\bibnamefont {Lapointe}}, \bibinfo
  {author} {\bibfnamefont {P.}~\bibnamefont {Grigorev}}, \bibinfo {author}
  {\bibfnamefont {T.~D.}\ \bibnamefont {Swinburne}}, \bibinfo {author}
  {\bibfnamefont {J.~R.}\ \bibnamefont {Kermode}}, \bibinfo {author}
  {\bibfnamefont {L.}~\bibnamefont {Ventelon}}, \bibinfo {author}
  {\bibfnamefont {J.}~\bibnamefont {Baima}}, \ and\ \bibinfo {author}
  {\bibfnamefont {M.-C.}\ \bibnamefont {Marinica}},\ }\href {\doibase
  10.1103/PhysRevMaterials.5.103803} {\bibfield  {journal} {\bibinfo  {journal}
  {Phys. Rev. Materials}\ }\textbf {\bibinfo {volume} {5}},\ \bibinfo {pages}
  {103803} (\bibinfo {year} {2021}{\natexlab{a}})}\BibitemShut {NoStop}%
\bibitem [{\citenamefont {Allen}\ \emph {et~al.}(2021)\citenamefont {Allen},
  \citenamefont {Dusson}, \citenamefont {Ortner},\ and\ \citenamefont
  {Cs{\'a}nyi}}]{allen2021atomic}%
  \BibitemOpen
  \bibfield  {author} {\bibinfo {author} {\bibfnamefont {A.~E.}\ \bibnamefont
  {Allen}}, \bibinfo {author} {\bibfnamefont {G.}~\bibnamefont {Dusson}},
  \bibinfo {author} {\bibfnamefont {C.}~\bibnamefont {Ortner}}, \ and\ \bibinfo
  {author} {\bibfnamefont {G.}~\bibnamefont {Cs{\'a}nyi}},\ }\href@noop {}
  {\bibfield  {journal} {\bibinfo  {journal} {Machine Learning: Science and
  Technology}\ }\textbf {\bibinfo {volume} {2}},\ \bibinfo {pages} {025017}
  (\bibinfo {year} {2021})}\BibitemShut {NoStop}%
\bibitem [{\citenamefont {Pun}\ \emph {et~al.}(2019)\citenamefont {Pun},
  \citenamefont {Batra}, \citenamefont {Ramprasad},\ and\ \citenamefont
  {Mishin}}]{pun2019physically}%
  \BibitemOpen
  \bibfield  {author} {\bibinfo {author} {\bibfnamefont {G.~P.}\ \bibnamefont
  {Pun}}, \bibinfo {author} {\bibfnamefont {R.}~\bibnamefont {Batra}}, \bibinfo
  {author} {\bibfnamefont {R.}~\bibnamefont {Ramprasad}}, \ and\ \bibinfo
  {author} {\bibfnamefont {Y.}~\bibnamefont {Mishin}},\ }\href@noop {}
  {\bibfield  {journal} {\bibinfo  {journal} {Nature communications}\ }\textbf
  {\bibinfo {volume} {10}},\ \bibinfo {pages} {1} (\bibinfo {year}
  {2019})}\BibitemShut {NoStop}%
\bibitem [{\citenamefont {Chmiela}\ \emph {et~al.}(2018)\citenamefont
  {Chmiela}, \citenamefont {Sauceda}, \citenamefont {M\"uller},\ and\
  \citenamefont {Tkatchenko}}]{Chmiela_towards_2018}%
  \BibitemOpen
  \bibfield  {author} {\bibinfo {author} {\bibfnamefont {S.}~\bibnamefont
  {Chmiela}}, \bibinfo {author} {\bibfnamefont {H.~E.}\ \bibnamefont
  {Sauceda}}, \bibinfo {author} {\bibfnamefont {K.-R.}\ \bibnamefont
  {M\"uller}}, \ and\ \bibinfo {author} {\bibfnamefont {A.}~\bibnamefont
  {Tkatchenko}},\ }\href {\doibase 10.1038/s41467-018-06169-2} {\bibfield
  {journal} {\bibinfo  {journal} {Nat. Commun.}\ }\textbf {\bibinfo {volume}
  {9}},\ \bibinfo {pages} {1} (\bibinfo {year} {2018})}\BibitemShut {NoStop}%
\bibitem [{\citenamefont {Lysogorskiy}\ \emph {et~al.}(2021)\citenamefont
  {Lysogorskiy}, \citenamefont {van~der Oord}, \citenamefont {Bochkarev},
  \citenamefont {Menon}, \citenamefont {Rinaldi}, \citenamefont
  {Hammerschmidt}, \citenamefont {Mrovec}, \citenamefont {Thompson},
  \citenamefont {Cs{\'a}nyi}, \citenamefont {Ortner} \emph
  {et~al.}}]{lysogorskiy2021performant}%
  \BibitemOpen
  \bibfield  {author} {\bibinfo {author} {\bibfnamefont {Y.}~\bibnamefont
  {Lysogorskiy}}, \bibinfo {author} {\bibfnamefont {C.}~\bibnamefont {van~der
  Oord}}, \bibinfo {author} {\bibfnamefont {A.}~\bibnamefont {Bochkarev}},
  \bibinfo {author} {\bibfnamefont {S.}~\bibnamefont {Menon}}, \bibinfo
  {author} {\bibfnamefont {M.}~\bibnamefont {Rinaldi}}, \bibinfo {author}
  {\bibfnamefont {T.}~\bibnamefont {Hammerschmidt}}, \bibinfo {author}
  {\bibfnamefont {M.}~\bibnamefont {Mrovec}}, \bibinfo {author} {\bibfnamefont
  {A.}~\bibnamefont {Thompson}}, \bibinfo {author} {\bibfnamefont
  {G.}~\bibnamefont {Cs{\'a}nyi}}, \bibinfo {author} {\bibfnamefont
  {C.}~\bibnamefont {Ortner}},  \emph {et~al.},\ }\href@noop {} {\bibfield
  {journal} {\bibinfo  {journal} {npj Computational Materials}\ }\textbf
  {\bibinfo {volume} {7}},\ \bibinfo {pages} {1} (\bibinfo {year}
  {2021})}\BibitemShut {NoStop}%
\bibitem [{\citenamefont {Drautz}(2019)}]{drautz_atomic_2019}%
  \BibitemOpen
  \bibfield  {author} {\bibinfo {author} {\bibfnamefont {R.}~\bibnamefont
  {Drautz}},\ }\href {\doibase 10.1103/PhysRevB.99.014104} {\bibfield
  {journal} {\bibinfo  {journal} {Phys. Rev. B}\ }\textbf {\bibinfo {volume}
  {99}},\ \bibinfo {pages} {014104} (\bibinfo {year} {2019})}\BibitemShut
  {NoStop}%
\bibitem [{\citenamefont {Drautz}(2020)}]{drautz_atomic_2020}%
  \BibitemOpen
  \bibfield  {author} {\bibinfo {author} {\bibfnamefont {R.}~\bibnamefont
  {Drautz}},\ }\href {\doibase 10.1103/PhysRevB.102.024104} {\bibfield
  {journal} {\bibinfo  {journal} {Phys. Rev. B}\ }\textbf {\bibinfo {volume}
  {102}},\ \bibinfo {pages} {024104} (\bibinfo {year} {2020})}\BibitemShut
  {NoStop}%
\bibitem [{\citenamefont {Peierls}(1940)}]{peierls1940size}%
  \BibitemOpen
  \bibfield  {author} {\bibinfo {author} {\bibfnamefont {R.}~\bibnamefont
  {Peierls}},\ }\href@noop {} {\bibfield  {journal} {\bibinfo  {journal}
  {Proceedings of the Physical Society (1926-1948)}\ }\textbf {\bibinfo
  {volume} {52}},\ \bibinfo {pages} {34} (\bibinfo {year} {1940})}\BibitemShut
  {NoStop}%
\bibitem [{\citenamefont {Bernstein}\ \emph {et~al.}(2019)\citenamefont
  {Bernstein}, \citenamefont {Cs\'anyi},\ and\ \citenamefont
  {Deringer}}]{Bernstein_novo_2019}%
  \BibitemOpen
  \bibfield  {author} {\bibinfo {author} {\bibfnamefont {N.}~\bibnamefont
  {Bernstein}}, \bibinfo {author} {\bibfnamefont {G.}~\bibnamefont {Cs\'anyi}},
  \ and\ \bibinfo {author} {\bibfnamefont {V.~L.}\ \bibnamefont {Deringer}},\
  }\href {\doibase 10.1038/s41524-019-0236-6} {\bibfield  {journal} {\bibinfo
  {journal} {npj Computat. Mater.}\ }\textbf {\bibinfo {volume} {5}},\ \bibinfo
  {pages} {99} (\bibinfo {year} {2019})}\BibitemShut {NoStop}%
\bibitem [{\citenamefont {Hodapp}\ and\ \citenamefont
  {Shapeev}(2020)}]{Hodapp2020}%
  \BibitemOpen
  \bibfield  {author} {\bibinfo {author} {\bibfnamefont {M.}~\bibnamefont
  {Hodapp}}\ and\ \bibinfo {author} {\bibfnamefont {A.}~\bibnamefont
  {Shapeev}},\ }\href {\doibase 10.1088/2632-2153/aba373} {\bibfield  {journal}
  {\bibinfo  {journal} {Machine Learning: Science and Technology}\ }\textbf
  {\bibinfo {volume} {1}},\ \bibinfo {pages} {45005} (\bibinfo {year}
  {2020})}\BibitemShut {NoStop}%
\bibitem [{\citenamefont {Vandermause}\ \emph {et~al.}(2020)\citenamefont
  {Vandermause}, \citenamefont {Torrisi}, \citenamefont {Batzner},
  \citenamefont {Xie}, \citenamefont {Sun}, \citenamefont {Kolpak},\ and\
  \citenamefont {Kozinsky}}]{vandermause2020fly}%
  \BibitemOpen
  \bibfield  {author} {\bibinfo {author} {\bibfnamefont {J.}~\bibnamefont
  {Vandermause}}, \bibinfo {author} {\bibfnamefont {S.~B.}\ \bibnamefont
  {Torrisi}}, \bibinfo {author} {\bibfnamefont {S.}~\bibnamefont {Batzner}},
  \bibinfo {author} {\bibfnamefont {Y.}~\bibnamefont {Xie}}, \bibinfo {author}
  {\bibfnamefont {L.}~\bibnamefont {Sun}}, \bibinfo {author} {\bibfnamefont
  {A.~M.}\ \bibnamefont {Kolpak}}, \ and\ \bibinfo {author} {\bibfnamefont
  {B.}~\bibnamefont {Kozinsky}},\ }\href@noop {} {\bibfield  {journal}
  {\bibinfo  {journal} {npj Computational Materials}\ }\textbf {\bibinfo
  {volume} {6}},\ \bibinfo {pages} {1} (\bibinfo {year} {2020})}\BibitemShut
  {NoStop}%
\bibitem [{\citenamefont {Rasmussen}(2004)}]{Book_Rasmussen}%
  \BibitemOpen
  \bibfield  {author} {\bibinfo {author} {\bibfnamefont {C.~E.}\ \bibnamefont
  {Rasmussen}},\ }\href {https://doi.org/10.1007/978-3-540-28650-9_4} {\emph
  {\bibinfo {title} {Gaussian Processes in Machine Learning}}}\ (\bibinfo
  {publisher} {Springer, Berlin, Heidelberg},\ \bibinfo {year}
  {2004})\BibitemShut {NoStop}%
\bibitem [{\citenamefont {Bart\'ok}\ \emph {et~al.}(2017)\citenamefont
  {Bart\'ok}, \citenamefont {De}, \citenamefont {Poelking}, \citenamefont
  {Bernstein}, \citenamefont {Kermode}, \citenamefont {Cs\'anyi},\ and\
  \citenamefont {Ceriotti}}]{Bartok_machine_2017}%
  \BibitemOpen
  \bibfield  {author} {\bibinfo {author} {\bibfnamefont {A.~P.}\ \bibnamefont
  {Bart\'ok}}, \bibinfo {author} {\bibfnamefont {S.}~\bibnamefont {De}},
  \bibinfo {author} {\bibfnamefont {C.}~\bibnamefont {Poelking}}, \bibinfo
  {author} {\bibfnamefont {N.}~\bibnamefont {Bernstein}}, \bibinfo {author}
  {\bibfnamefont {J.~R.}\ \bibnamefont {Kermode}}, \bibinfo {author}
  {\bibfnamefont {G.}~\bibnamefont {Cs\'anyi}}, \ and\ \bibinfo {author}
  {\bibfnamefont {M.}~\bibnamefont {Ceriotti}},\ }\href {\doibase
  10.1126/sciadv.1701816} {\bibfield  {journal} {\bibinfo  {journal} {Sci.
  Adv.}\ }\textbf {\bibinfo {volume} {3}},\ \bibinfo {pages} {e1701816}
  (\bibinfo {year} {2017})}\BibitemShut {NoStop}%
\bibitem [{\citenamefont {Goryaeva}\ \emph {et~al.}(2020)\citenamefont
  {Goryaeva}, \citenamefont {Lapointe}, \citenamefont {Dai}, \citenamefont
  {D{\'e}r{\`e}s}, \citenamefont {Maillet},\ and\ \citenamefont
  {Marinica}}]{goryaeva2020}%
  \BibitemOpen
  \bibfield  {author} {\bibinfo {author} {\bibfnamefont {A.~M.}\ \bibnamefont
  {Goryaeva}}, \bibinfo {author} {\bibfnamefont {C.}~\bibnamefont {Lapointe}},
  \bibinfo {author} {\bibfnamefont {C.}~\bibnamefont {Dai}}, \bibinfo {author}
  {\bibfnamefont {J.}~\bibnamefont {D{\'e}r{\`e}s}}, \bibinfo {author}
  {\bibfnamefont {J.-B.}\ \bibnamefont {Maillet}}, \ and\ \bibinfo {author}
  {\bibfnamefont {M.-C.}\ \bibnamefont {Marinica}},\ }\href@noop {} {\bibfield
  {journal} {\bibinfo  {journal} {Nature communications}\ }\textbf {\bibinfo
  {volume} {11}},\ \bibinfo {pages} {1} (\bibinfo {year} {2020})}\BibitemShut
  {NoStop}%
\bibitem [{\citenamefont {Larsen}\ \emph {et~al.}(2017)\citenamefont {Larsen},
  \citenamefont {Mortensen}, \citenamefont {Blomqvist}, \citenamefont
  {Castelli}, \citenamefont {Christensen}, \citenamefont {Du{\l}ak},
  \citenamefont {Friis}, \citenamefont {Groves}, \citenamefont {Hammer},
  \citenamefont {Hargus}, \citenamefont {Hermes}, \citenamefont {Jennings},
  \citenamefont {Jensen}, \citenamefont {Kermode}, \citenamefont {Kitchin},
  \citenamefont {Kolsbjerg}, \citenamefont {Kubal}, \citenamefont {Kaasbjerg},
  \citenamefont {Lysgaard}, \citenamefont {Maronsson}, \citenamefont {Maxson},
  \citenamefont {Olsen}, \citenamefont {Pastewka}, \citenamefont {Peterson},
  \citenamefont {Rostgaard}, \citenamefont {Schi{\o}tz}, \citenamefont
  {Sch{\"u}tt}, \citenamefont {Strange}, \citenamefont {Thygesen},
  \citenamefont {Vegge}, \citenamefont {Vilhelmsen}, \citenamefont {Walter},
  \citenamefont {Zeng},\ and\ \citenamefont {Jacobsen}}]{Hjorth_Larsen_2017}%
  \BibitemOpen
  \bibfield  {author} {\bibinfo {author} {\bibfnamefont {A.~H.}\ \bibnamefont
  {Larsen}}, \bibinfo {author} {\bibfnamefont {J.~J.}\ \bibnamefont
  {Mortensen}}, \bibinfo {author} {\bibfnamefont {J.}~\bibnamefont
  {Blomqvist}}, \bibinfo {author} {\bibfnamefont {I.~E.}\ \bibnamefont
  {Castelli}}, \bibinfo {author} {\bibfnamefont {R.}~\bibnamefont
  {Christensen}}, \bibinfo {author} {\bibfnamefont {M.}~\bibnamefont
  {Du{\l}ak}}, \bibinfo {author} {\bibfnamefont {J.}~\bibnamefont {Friis}},
  \bibinfo {author} {\bibfnamefont {M.~N.}\ \bibnamefont {Groves}}, \bibinfo
  {author} {\bibfnamefont {B.}~\bibnamefont {Hammer}}, \bibinfo {author}
  {\bibfnamefont {C.}~\bibnamefont {Hargus}}, \bibinfo {author} {\bibfnamefont
  {E.~D.}\ \bibnamefont {Hermes}}, \bibinfo {author} {\bibfnamefont {P.~C.}\
  \bibnamefont {Jennings}}, \bibinfo {author} {\bibfnamefont {P.~B.}\
  \bibnamefont {Jensen}}, \bibinfo {author} {\bibfnamefont {J.}~\bibnamefont
  {Kermode}}, \bibinfo {author} {\bibfnamefont {J.~R.}\ \bibnamefont
  {Kitchin}}, \bibinfo {author} {\bibfnamefont {E.~L.}\ \bibnamefont
  {Kolsbjerg}}, \bibinfo {author} {\bibfnamefont {J.}~\bibnamefont {Kubal}},
  \bibinfo {author} {\bibfnamefont {K.}~\bibnamefont {Kaasbjerg}}, \bibinfo
  {author} {\bibfnamefont {S.}~\bibnamefont {Lysgaard}}, \bibinfo {author}
  {\bibfnamefont {J.~B.}\ \bibnamefont {Maronsson}}, \bibinfo {author}
  {\bibfnamefont {T.}~\bibnamefont {Maxson}}, \bibinfo {author} {\bibfnamefont
  {T.}~\bibnamefont {Olsen}}, \bibinfo {author} {\bibfnamefont
  {L.}~\bibnamefont {Pastewka}}, \bibinfo {author} {\bibfnamefont
  {A.}~\bibnamefont {Peterson}}, \bibinfo {author} {\bibfnamefont
  {C.}~\bibnamefont {Rostgaard}}, \bibinfo {author} {\bibfnamefont
  {J.}~\bibnamefont {Schi{\o}tz}}, \bibinfo {author} {\bibfnamefont
  {O.}~\bibnamefont {Sch{\"u}tt}}, \bibinfo {author} {\bibfnamefont
  {M.}~\bibnamefont {Strange}}, \bibinfo {author} {\bibfnamefont {K.~S.}\
  \bibnamefont {Thygesen}}, \bibinfo {author} {\bibfnamefont {T.}~\bibnamefont
  {Vegge}}, \bibinfo {author} {\bibfnamefont {L.}~\bibnamefont {Vilhelmsen}},
  \bibinfo {author} {\bibfnamefont {M.}~\bibnamefont {Walter}}, \bibinfo
  {author} {\bibfnamefont {Z.}~\bibnamefont {Zeng}}, \ and\ \bibinfo {author}
  {\bibfnamefont {K.~W.}\ \bibnamefont {Jacobsen}},\ }\href {\doibase
  10.1088/1361-648x/aa680e} {\bibfield  {journal} {\bibinfo  {journal} {Journal
  of Physics: Condensed Matter}\ }\textbf {\bibinfo {volume} {29}},\ \bibinfo
  {pages} {273002} (\bibinfo {year} {2017})}\BibitemShut {NoStop}%
\bibitem [{\citenamefont {Grigorev}\ and\ \citenamefont
  {Swinburne}(2021{\natexlab{a}})}]{code}%
  \BibitemOpen
  \bibfield  {author} {\bibinfo {author} {\bibfnamefont {P.}~\bibnamefont
  {Grigorev}}\ and\ \bibinfo {author} {\bibfnamefont {T.~D.}\ \bibnamefont
  {Swinburne}},\ }\href@noop {} {\enquote {\bibinfo {title} {{LML} constrained
  retraining package},}\ }\bibinfo {howpublished}
  {\url{https://github.com/marseille-matmol/LML-retrain}} (\bibinfo {year}
  {2021}{\natexlab{a}})\BibitemShut {NoStop}%
\bibitem [{\citenamefont {Goryaeva}\ \emph
  {et~al.}(2021{\natexlab{b}})\citenamefont {Goryaeva}, \citenamefont
  {Lapointe}, \citenamefont {Swinburne},\ and\ \citenamefont
  {Marinica}}]{milady}%
  \BibitemOpen
  \bibfield  {author} {\bibinfo {author} {\bibfnamefont {A.~M.}\ \bibnamefont
  {Goryaeva}}, \bibinfo {author} {\bibfnamefont {C.}~\bibnamefont {Lapointe}},
  \bibinfo {author} {\bibfnamefont {T.~D.}\ \bibnamefont {Swinburne}}, \ and\
  \bibinfo {author} {\bibfnamefont {M.-C.}\ \bibnamefont {Marinica}},\
  }\href@noop {} {\enquote {\bibinfo {title} {{Lammps-MiLaDy} package},}\
  }\bibinfo {howpublished} {\url{https://github.com/ai-atoms/Lammps-MiLaDy}}
  (\bibinfo {year} {2021}{\natexlab{b}})\BibitemShut {NoStop}%
\bibitem [{\citenamefont {Strang}(1993)}]{strang}%
  \BibitemOpen
  \bibfield  {author} {\bibinfo {author} {\bibfnamefont {G.}~\bibnamefont
  {Strang}},\ }\href@noop {} {\emph {\bibinfo {title} {Introduction to linear
  algebra}}},\ Vol.~\bibinfo {volume} {3}\ (\bibinfo  {publisher}
  {Wellesley-Cambridge Press Wellesley, MA},\ \bibinfo {year}
  {1993})\BibitemShut {NoStop}%
\bibitem [{\citenamefont {Weinberger}\ \emph {et~al.}(2013)\citenamefont
  {Weinberger}, \citenamefont {Boyce},\ and\ \citenamefont
  {Battaile}}]{Weinberger2013}%
  \BibitemOpen
  \bibfield  {author} {\bibinfo {author} {\bibfnamefont {C.~R.}\ \bibnamefont
  {Weinberger}}, \bibinfo {author} {\bibfnamefont {B.~L.}\ \bibnamefont
  {Boyce}}, \ and\ \bibinfo {author} {\bibfnamefont {C.~C.}\ \bibnamefont
  {Battaile}},\ }\href {\doibase 10.1179/1743280412Y.0000000015} {\bibfield
  {journal} {\bibinfo  {journal} {International Materials Reviews}\ }\textbf
  {\bibinfo {volume} {58}} (\bibinfo {year} {2013}),\
  10.1179/1743280412Y.0000000015}\BibitemShut {NoStop}%
\bibitem [{\citenamefont {Clouet}\ \emph {et~al.}(2021)\citenamefont {Clouet},
  \citenamefont {Bienvenu}, \citenamefont {Dezerald},\ and\ \citenamefont
  {Rodney}}]{Clouet2021}%
  \BibitemOpen
  \bibfield  {author} {\bibinfo {author} {\bibfnamefont {E.}~\bibnamefont
  {Clouet}}, \bibinfo {author} {\bibfnamefont {B.}~\bibnamefont {Bienvenu}},
  \bibinfo {author} {\bibfnamefont {L.}~\bibnamefont {Dezerald}}, \ and\
  \bibinfo {author} {\bibfnamefont {D.}~\bibnamefont {Rodney}},\ }\href
  {\doibase 10.5802/CRPHYS.75} {\bibfield  {journal} {\bibinfo  {journal}
  {Comptes Rendus Physique}\ }\textbf {\bibinfo {volume} {22}} (\bibinfo {year}
  {2021}),\ 10.5802/CRPHYS.75}\BibitemShut {NoStop}%
\bibitem [{\citenamefont {Bulatov}\ \emph {et~al.}(2006)\citenamefont
  {Bulatov}, \citenamefont {Hsiung}, \citenamefont {Tang}, \citenamefont
  {Arsenlis}, \citenamefont {Bartelt}, \citenamefont {Cai}, \citenamefont
  {Florando}, \citenamefont {Hiratani}, \citenamefont {Rhee}, \citenamefont
  {Hommes}, \citenamefont {Pierce},\ and\ \citenamefont {Rubia}}]{Bulatov2006}%
  \BibitemOpen
  \bibfield  {author} {\bibinfo {author} {\bibfnamefont {V.~V.}\ \bibnamefont
  {Bulatov}}, \bibinfo {author} {\bibfnamefont {L.~L.}\ \bibnamefont {Hsiung}},
  \bibinfo {author} {\bibfnamefont {M.}~\bibnamefont {Tang}}, \bibinfo {author}
  {\bibfnamefont {A.}~\bibnamefont {Arsenlis}}, \bibinfo {author}
  {\bibfnamefont {M.~C.}\ \bibnamefont {Bartelt}}, \bibinfo {author}
  {\bibfnamefont {W.}~\bibnamefont {Cai}}, \bibinfo {author} {\bibfnamefont
  {J.~N.}\ \bibnamefont {Florando}}, \bibinfo {author} {\bibfnamefont
  {M.}~\bibnamefont {Hiratani}}, \bibinfo {author} {\bibfnamefont
  {M.}~\bibnamefont {Rhee}}, \bibinfo {author} {\bibfnamefont {G.}~\bibnamefont
  {Hommes}}, \bibinfo {author} {\bibfnamefont {T.~G.}\ \bibnamefont {Pierce}},
  \ and\ \bibinfo {author} {\bibfnamefont {T.~D. D.~L.}\ \bibnamefont
  {Rubia}},\ }\href {\doibase 10.1038/nature04658} {\bibfield  {journal}
  {\bibinfo  {journal} {Nature}\ }\textbf {\bibinfo {volume} {440}} (\bibinfo
  {year} {2006}),\ 10.1038/nature04658}\BibitemShut {NoStop}%
\bibitem [{\citenamefont {Guo}\ \emph {et~al.}(2019)\citenamefont {Guo},
  \citenamefont {Ge}, \citenamefont {Yuan}, \citenamefont {Cheng},
  \citenamefont {Wang}, \citenamefont {Zhang},\ and\ \citenamefont
  {Lu}}]{Guo2019}%
  \BibitemOpen
  \bibfield  {author} {\bibinfo {author} {\bibfnamefont {W.}~\bibnamefont
  {Guo}}, \bibinfo {author} {\bibfnamefont {L.}~\bibnamefont {Ge}}, \bibinfo
  {author} {\bibfnamefont {Y.}~\bibnamefont {Yuan}}, \bibinfo {author}
  {\bibfnamefont {L.}~\bibnamefont {Cheng}}, \bibinfo {author} {\bibfnamefont
  {S.}~\bibnamefont {Wang}}, \bibinfo {author} {\bibfnamefont {X.}~\bibnamefont
  {Zhang}}, \ and\ \bibinfo {author} {\bibfnamefont {G.~H.}\ \bibnamefont
  {Lu}},\ }\href {\doibase 10.1088/1741-4326/aaf32e} {\bibfield  {journal}
  {\bibinfo  {journal} {Nuclear Fusion}\ }\textbf {\bibinfo {volume} {59}}
  (\bibinfo {year} {2019}),\ 10.1088/1741-4326/aaf32e}\BibitemShut {NoStop}%
\bibitem [{\citenamefont {Bertin}\ \emph
  {et~al.}(2021{\natexlab{a}})\citenamefont {Bertin}, \citenamefont {Cai},
  \citenamefont {Aubry},\ and\ \citenamefont {Bulatov}}]{Bertin2021}%
  \BibitemOpen
  \bibfield  {author} {\bibinfo {author} {\bibfnamefont {N.}~\bibnamefont
  {Bertin}}, \bibinfo {author} {\bibfnamefont {W.}~\bibnamefont {Cai}},
  \bibinfo {author} {\bibfnamefont {S.}~\bibnamefont {Aubry}}, \ and\ \bibinfo
  {author} {\bibfnamefont {V.~V.}\ \bibnamefont {Bulatov}},\ }\href {\doibase
  10.1103/PhysRevMaterials.5.025002} {\bibfield  {journal} {\bibinfo  {journal}
  {Physical Review Materials}\ }\textbf {\bibinfo {volume} {5}} (\bibinfo
  {year} {2021}{\natexlab{a}}),\ 10.1103/PhysRevMaterials.5.025002}\BibitemShut
  {NoStop}%
\bibitem [{\citenamefont {Marinica}\ \emph {et~al.}(2013)\citenamefont
  {Marinica}, \citenamefont {Ventelon}, \citenamefont {Gilbert}, \citenamefont
  {Proville}, \citenamefont {Dudarev}, \citenamefont {Marian}, \citenamefont
  {Bencteux},\ and\ \citenamefont {Willaime}}]{Marinica2013}%
  \BibitemOpen
  \bibfield  {author} {\bibinfo {author} {\bibfnamefont {M.~C.}\ \bibnamefont
  {Marinica}}, \bibinfo {author} {\bibfnamefont {L.}~\bibnamefont {Ventelon}},
  \bibinfo {author} {\bibfnamefont {M.~R.}\ \bibnamefont {Gilbert}}, \bibinfo
  {author} {\bibfnamefont {L.}~\bibnamefont {Proville}}, \bibinfo {author}
  {\bibfnamefont {S.~L.}\ \bibnamefont {Dudarev}}, \bibinfo {author}
  {\bibfnamefont {J.}~\bibnamefont {Marian}}, \bibinfo {author} {\bibfnamefont
  {G.}~\bibnamefont {Bencteux}}, \ and\ \bibinfo {author} {\bibfnamefont
  {F.}~\bibnamefont {Willaime}},\ }\href {\doibase
  10.1088/0953-8984/25/39/395502} {\bibfield  {journal} {\bibinfo  {journal}
  {Journal of Physics Condensed Matter}\ }\textbf {\bibinfo {volume} {25}}
  (\bibinfo {year} {2013}),\ 10.1088/0953-8984/25/39/395502}\BibitemShut
  {NoStop}%
\bibitem [{\citenamefont {Grigorev}\ and\ \citenamefont
  {Swinburne}(2021{\natexlab{b}})}]{sm}%
  \BibitemOpen
  \bibfield  {author} {\bibinfo {author} {\bibfnamefont {P.}~\bibnamefont
  {Grigorev}}\ and\ \bibinfo {author} {\bibfnamefont {T.~D.}\ \bibnamefont
  {Swinburne}},\ }\href@noop {} {\enquote {\bibinfo {title} {Supplementary
  material},}\ } (\bibinfo {year} {2021}{\natexlab{b}})\BibitemShut {NoStop}%
\bibitem [{\citenamefont {Bertin}\ \emph
  {et~al.}(2021{\natexlab{b}})\citenamefont {Bertin}, \citenamefont {Cai},
  \citenamefont {Aubry},\ and\ \citenamefont {Bulatov}}]{bertin2021core}%
  \BibitemOpen
  \bibfield  {author} {\bibinfo {author} {\bibfnamefont {N.}~\bibnamefont
  {Bertin}}, \bibinfo {author} {\bibfnamefont {W.}~\bibnamefont {Cai}},
  \bibinfo {author} {\bibfnamefont {S.}~\bibnamefont {Aubry}}, \ and\ \bibinfo
  {author} {\bibfnamefont {V.}~\bibnamefont {Bulatov}},\ }\href@noop {}
  {\bibfield  {journal} {\bibinfo  {journal} {Physical Review Materials}\
  }\textbf {\bibinfo {volume} {5}},\ \bibinfo {pages} {025002} (\bibinfo {year}
  {2021}{\natexlab{b}})}\BibitemShut {NoStop}%
\bibitem [{\citenamefont {Vitek}(1974)}]{Vitek1974}%
  \BibitemOpen
  \bibfield  {author} {\bibinfo {author} {\bibfnamefont {V.}~\bibnamefont
  {Vitek}},\ }\href@noop {} {\bibfield  {journal} {\bibinfo  {journal} {Crystal
  Lattice Defects}\ }\textbf {\bibinfo {volume} {5}},\ \bibinfo {pages} {pp. 1}
  (\bibinfo {year} {1974})}\BibitemShut {NoStop}%
\bibitem [{\citenamefont {Takeuchi}(1979)}]{Takeuchi1979}%
  \BibitemOpen
  \bibfield  {author} {\bibinfo {author} {\bibfnamefont {S.}~\bibnamefont
  {Takeuchi}},\ }\href@noop {} {\bibfield  {journal} {\bibinfo  {journal}
  {Philosophical Magazine A}\ }\textbf {\bibinfo {volume} {39}},\ \bibinfo
  {pages} {661} (\bibinfo {year} {1979})}\BibitemShut {NoStop}%
\bibitem [{\citenamefont {Cai}\ \emph {et~al.}(2004)\citenamefont {Cai},
  \citenamefont {Bulatov}, \citenamefont {Chang}, \citenamefont {Li},\ and\
  \citenamefont {Yip}}]{CAI_2004}%
  \BibitemOpen
  \bibfield  {author} {\bibinfo {author} {\bibfnamefont {W.}~\bibnamefont
  {Cai}}, \bibinfo {author} {\bibfnamefont {V.~V.}\ \bibnamefont {Bulatov}},
  \bibinfo {author} {\bibfnamefont {J.}~\bibnamefont {Chang}}, \bibinfo
  {author} {\bibfnamefont {J.}~\bibnamefont {Li}}, \ and\ \bibinfo {author}
  {\bibfnamefont {S.}~\bibnamefont {Yip}},\ }in\ \href {\doibase
  https://doi.org/10.1016/S1572-4859(05)80003-8} {\emph {\bibinfo {booktitle}
  {Dislocations in Solids}}},\ \bibinfo {series} {Dislocations in Solids},
  Vol.~\bibinfo {volume} {12},\ \bibinfo {editor} {edited by\ \bibinfo {editor}
  {\bibfnamefont {F.}~\bibnamefont {Nabarro}}\ and\ \bibinfo {editor}
  {\bibfnamefont {J.}~\bibnamefont {Hirth}}}\ (\bibinfo  {publisher}
  {Elsevier},\ \bibinfo {year} {2004})\ pp.\ \bibinfo {pages} {1 --
  80}\BibitemShut {NoStop}%
\bibitem [{\citenamefont {Ventelon}\ \emph {et~al.}(2013)\citenamefont
  {Ventelon}, \citenamefont {Willaime}, \citenamefont {Clouet},\ and\
  \citenamefont {Rodney}}]{Ventelon2013}%
  \BibitemOpen
  \bibfield  {author} {\bibinfo {author} {\bibfnamefont {L.}~\bibnamefont
  {Ventelon}}, \bibinfo {author} {\bibfnamefont {F.}~\bibnamefont {Willaime}},
  \bibinfo {author} {\bibfnamefont {E.}~\bibnamefont {Clouet}}, \ and\ \bibinfo
  {author} {\bibfnamefont {D.}~\bibnamefont {Rodney}},\ }\href {\doibase
  https://doi.org/10.1016/j.actamat.2013.03.012} {\bibfield  {journal}
  {\bibinfo  {journal} {Acta Materialia}\ }\textbf {\bibinfo {volume} {61}},\
  \bibinfo {pages} {3973 } (\bibinfo {year} {2013})}\BibitemShut {NoStop}%
\bibitem [{\citenamefont {Kermode}\ \emph {et~al.}(2020)\citenamefont
  {Kermode}, \citenamefont {Pastewka},\ and\ \citenamefont
  {Grigorev}}]{matscipy}%
  \BibitemOpen
  \bibfield  {author} {\bibinfo {author} {\bibfnamefont {J.~R.}\ \bibnamefont
  {Kermode}}, \bibinfo {author} {\bibfnamefont {L.}~\bibnamefont {Pastewka}}, \
  and\ \bibinfo {author} {\bibfnamefont {P.}~\bibnamefont {Grigorev}},\
  }\href@noop {} {\enquote {\bibinfo {title} {matscipy: generic python
  materials science toolkit},}\ }\bibinfo {howpublished}
  {\url{https://github.com/libAtoms/matscipy}} (\bibinfo {year}
  {2020})\BibitemShut {NoStop}%
\bibitem [{\citenamefont {Bitzek}\ \emph {et~al.}(2006)\citenamefont {Bitzek},
  \citenamefont {Koskinen}, \citenamefont {G\"ahler}, \citenamefont {Moseler},\
  and\ \citenamefont {Gumbsch}}]{FIRE2006}%
  \BibitemOpen
  \bibfield  {author} {\bibinfo {author} {\bibfnamefont {E.}~\bibnamefont
  {Bitzek}}, \bibinfo {author} {\bibfnamefont {P.}~\bibnamefont {Koskinen}},
  \bibinfo {author} {\bibfnamefont {F.}~\bibnamefont {G\"ahler}}, \bibinfo
  {author} {\bibfnamefont {M.}~\bibnamefont {Moseler}}, \ and\ \bibinfo
  {author} {\bibfnamefont {P.}~\bibnamefont {Gumbsch}},\ }\href {\doibase
  10.1103/PhysRevLett.97.170201} {\bibfield  {journal} {\bibinfo  {journal}
  {Phys. Rev. Lett.}\ }\textbf {\bibinfo {volume} {97}},\ \bibinfo {pages}
  {170201} (\bibinfo {year} {2006})}\BibitemShut {NoStop}%
\bibitem [{\citenamefont {Packwood}\ \emph {et~al.}(2016)\citenamefont
  {Packwood}, \citenamefont {Kermode}, \citenamefont {Mones}, \citenamefont
  {Bernstein}, \citenamefont {Woolley}, \citenamefont {Gould}, \citenamefont
  {Ortner},\ and\ \citenamefont {Cs{\'a}nyi}}]{Packwood2016}%
  \BibitemOpen
  \bibfield  {author} {\bibinfo {author} {\bibfnamefont {D.}~\bibnamefont
  {Packwood}}, \bibinfo {author} {\bibfnamefont {J.}~\bibnamefont {Kermode}},
  \bibinfo {author} {\bibfnamefont {L.}~\bibnamefont {Mones}}, \bibinfo
  {author} {\bibfnamefont {N.}~\bibnamefont {Bernstein}}, \bibinfo {author}
  {\bibfnamefont {J.}~\bibnamefont {Woolley}}, \bibinfo {author} {\bibfnamefont
  {N.}~\bibnamefont {Gould}}, \bibinfo {author} {\bibfnamefont
  {C.}~\bibnamefont {Ortner}}, \ and\ \bibinfo {author} {\bibfnamefont
  {G.}~\bibnamefont {Cs{\'a}nyi}},\ }\href {\doibase 10.1063/1.4947024}
  {\bibfield  {journal} {\bibinfo  {journal} {J. Chem. Phys.}\ }\textbf
  {\bibinfo {volume} {144}},\ \bibinfo {pages} {164109} (\bibinfo {year}
  {2016})}\BibitemShut {NoStop}%
\bibitem [{\citenamefont {Mones}\ \emph {et~al.}(2018)\citenamefont {Mones},
  \citenamefont {Ortner},\ and\ \citenamefont
  {Cs{\'a}nyi}}]{mones2018preconditioners}%
  \BibitemOpen
  \bibfield  {author} {\bibinfo {author} {\bibfnamefont {L.}~\bibnamefont
  {Mones}}, \bibinfo {author} {\bibfnamefont {C.}~\bibnamefont {Ortner}}, \
  and\ \bibinfo {author} {\bibfnamefont {G.}~\bibnamefont {Cs{\'a}nyi}},\
  }\href@noop {} {\bibfield  {journal} {\bibinfo  {journal} {Scientific
  reports}\ }\textbf {\bibinfo {volume} {8}},\ \bibinfo {pages} {1} (\bibinfo
  {year} {2018})}\BibitemShut {NoStop}%
\bibitem [{\citenamefont {Makri}\ \emph {et~al.}(2019)\citenamefont {Makri},
  \citenamefont {Ortner},\ and\ \citenamefont {Kermode}}]{Makri2019}%
  \BibitemOpen
  \bibfield  {author} {\bibinfo {author} {\bibfnamefont {S.}~\bibnamefont
  {Makri}}, \bibinfo {author} {\bibfnamefont {C.}~\bibnamefont {Ortner}}, \
  and\ \bibinfo {author} {\bibfnamefont {J.~R.}\ \bibnamefont {Kermode}},\
  }\href {\doibase 10.1063/1.5064465} {\bibfield  {journal} {\bibinfo
  {journal} {Journal of Chemical Physics}\ }\textbf {\bibinfo {volume} {150}}
  (\bibinfo {year} {2019}),\ 10.1063/1.5064465}\BibitemShut {NoStop}%
\bibitem [{\citenamefont {Kresse}\ and\ \citenamefont
  {Furthm\"uller}(1996)}]{VASP}%
  \BibitemOpen
  \bibfield  {author} {\bibinfo {author} {\bibfnamefont {G.}~\bibnamefont
  {Kresse}}\ and\ \bibinfo {author} {\bibfnamefont {J.}~\bibnamefont
  {Furthm\"uller}},\ }\href {\doibase 10.1103/PhysRevB.54.11169} {\bibfield
  {journal} {\bibinfo  {journal} {Phys. Rev. B}\ }\textbf {\bibinfo {volume}
  {54}},\ \bibinfo {pages} {11169} (\bibinfo {year} {1996})}\BibitemShut
  {NoStop}%
\bibitem [{\citenamefont {Perdew}\ \emph {et~al.}(1996)\citenamefont {Perdew},
  \citenamefont {Burke},\ and\ \citenamefont {Ernzerhof}}]{PBE}%
  \BibitemOpen
  \bibfield  {author} {\bibinfo {author} {\bibfnamefont {J.~P.}\ \bibnamefont
  {Perdew}}, \bibinfo {author} {\bibfnamefont {K.}~\bibnamefont {Burke}}, \
  and\ \bibinfo {author} {\bibfnamefont {M.}~\bibnamefont {Ernzerhof}},\ }\href
  {\doibase 10.1103/PhysRevLett.77.3865} {\bibfield  {journal} {\bibinfo
  {journal} {Phys. Rev. Lett.}\ }\textbf {\bibinfo {volume} {77}},\ \bibinfo
  {pages} {3865} (\bibinfo {year} {1996})}\BibitemShut {NoStop}%
\end{thebibliography}%

\end{document}